\documentclass[10pt,preprint]{aastex}

\newcommand{\etal}{et~al.}
\newcommand{\ie}{{\it i.e.}}
\newcommand{\eg}{{\it e.g.}}
\newcommand{\cf}{{\rm cf.}}
\newcommand{\teff}{$T_{\rm eff}$}
\newcommand{\lbol}{$L_{\rm bol}$}
\newcommand{\msun}{$M_{\sun}$}
\newcommand{\ic}{$I_{\rm C}$}

\newcommand{\ik}{$I_{\rm C}-K_{\rm s}$}
\newcommand{\vi}{$V-I_{\rm C}$}
\newcommand{\evi}{E($V-I_{\rm C}$)}

\newcommand{\ai}{$A_I$}

\newcommand{\mdot}{$\dot{M}$}
\newcommand{\vsini}{$v \sin i$}

\received{}
\revised{}
\accepted{}
\shorttitle{Stellar Rotation and Radius}
\shortauthors{Rebull \etal}

\begin{document}

\title{The Early Angular Momentum History of Low Mass Stars:
	Evidence for a Regulation Mechanism}
\author{L.\ M.\ Rebull\altaffilmark{1,2}, S.\ C.\ Wolff\altaffilmark{3},
      S.\ E.\ Strom\altaffilmark{3}, R.\ B.\ Makidon\altaffilmark{4}}

\altaffiltext{1}{National Research Council Resident Research Associate,
	Jet Propulsion Laboratory, California Institute of Technology, M/S
	169-506, 4800 Oak Grove Drive, Pasadena, CA 91109
	(luisa.rebull@jpl.nasa.gov)}
\altaffiltext{2}{Currently Staff Scientist at SIRTF Science Center,
Caltech M/S 220-6, 1200 E. California Blvd., Pasadena, CA 91125}
\altaffiltext{3}{NOAO, 950 N.\ Cherry Ave, Tucson, AZ 87526}
\altaffiltext{4}{STScI, 3700 San Martin Dr, Baltimore, MD 21218}

\begin{abstract}

We examine the early angular momentum history of stars in young clusters
via 197 photometric periods in fields flanking the Orion Nebula Cluster
(ONC), 81 photometric periods in NGC 2264, and 202 measurements of $v \sin
i$ in the ONC itself.  We show that PMS stars spanning an age range from
0.1 to 3 Myr do not appear to conserve stellar angular momentum as they
evolve down their convective tracks, but instead preserve the same range
of periods even though they have contracted by about a factor of three.
This result seems to require a mechanism that regulates the angular
velocities of young stars. We discuss several candidate mechanisms.  The
most plausible appears to be disk-locking, though most of our stars do not
have \ik\ excesses suggestive of disks.  However, a decisive test of this
hypothesis requires a more sensitive diagnostic than the \ik\ excesses used
here.
\end{abstract}

\keywords{stars: pre-main sequence --- stars: rotation}

\section{Introduction}

Observations over the past decade have established that pre-main sequence
(PMS) stars accrete a substantial portion of their final mass via material
transported through circumstellar disks. However, the prediction that
accretion of high angular momentum disk material will cause the stars to
spin up during the accretion phase (\eg\ Durisen \etal\ 1989) is not borne
out by observations: most PMS stars have rotational velocities ($v$) of no
more than a few tens of km s$^{-1}$.  By contrast, the typical breakup
velocity at $\sim$1 Myr is $\sim$300 km~s$^{-1}$.

Explanations of the observed slow rotation typically invoke angular
momentum transfer either from the star to the surrounding accretion disk
(\eg\ K\"onigl 1991), or to a stellar wind originating at the boundary
between the disk and the stellar magnetosphere (\eg\ Shu \etal\ 2000).
Either mechanism requires that during the disk accretion phase, the
angular velocity of the star be  ``locked" to a period set by the
Keplerian angular velocity at or near the boundary between the stellar
magnetosphere and the accretion disk.  At the end of the accretion phase,
PMS stars should be unlocked from their disks, and free to spin up as they
contract toward the main sequence.

Considerable observational effort has been devoted to testing the
prediction that rotation periods are locked to a narrow range during the
accretion phase via comparison of the distribution of stellar periods or
rotation rates for samples of stars surrounded by disks with those that
lack disks.  Early studies suggested that stars surrounded by disks tend
to have significantly longer rotation periods than their diskless
counterparts (\eg\ Edwards \etal\ 1993, Choi \& Herbst 1996); more recent
results based on larger samples yield more ambiguous results (\eg\ Rebull
2001 [R01]; Stassun \etal\ 1999 [SMMV]; see also Herbst \etal\ 2000
[HRHC], Carpenter \etal\ 2001 [CHS], Herbst \etal\ 2001 [HBJM]). HRHC and
HBJM both find a bimodal distribution of periods for the higher-mass
($\gtrsim$0.25\msun) stars, whereas it is not bimodal for the lower mass
stars.  R01 looked for this dependence, but did not find it.   

Over the past few years, rotation periods $P$ or projected rotational
velocities $v \sin i$ have been measured for large samples of PMS stars.
These data sets are now large enough to to map changes in stellar angular
momentum ($J$) as stars of different masses evolve down their convective
tracks, and to understand the role that disk locking may play in determining
the angular momenta of young stars.

In this paper, we exploit these large datasets to examine the early
angular momentum history of PMS stars spanning ages from  $\sim$0.1 Myr to
$\sim$3 Myr and masses $\sim$0.2 to $\sim$2 \msun, both for stars
apparently surrounded by accretion disks and those that lack evidence of
such disks.

We first examine two samples of low mass PMS stars for which rotation
periods are derived from observations of spot-modulated light variations:
stars located in fields flanking the Orion Nebula Cluster and NGC 2264.
Together, stars comprising these samples span masses from $\sim$0.2$-$2
\msun\ and and nominal ages from $\sim$0.1$-$3 Myr.  At fixed \teff, radii
appear to span a range of about a factor of 3.  Absent any external
angular momentum loss mechanism (e.g., disk locking or spindown torques
exerted by stellar winds), contraction of these fully convective stars
over this radius range should result in a dramatic (9-fold) decrease in
period. Hence, our sample appears well-suited to quantifying changes in
period distributions among PMS stars arising as stars contract along
convective tracks, and to assessing the possible role of mechanisms that
could potentially regulate or mitigate such evolution-driven changes among
low mass stars.

Our approach will be to examine the relationship between observed
rotation period ($P$) and derived stellar radius ($R$) for  a large sample
of PMS stars -- both those which appear surrounded by circumstellar
accretion disks (as judged from near-IR excess emission), and those that
appear to lack such disks. We thus first summarize the observational
database for each sample: the derived periods as well as sample
limitations, along with the photometry and spectroscopy that enable
locating each star in the \lbol$-$\teff\ plane (thus yielding a radius)
and determining whether or not a star is surrounded by a circumstellar
accretion disk.  Particular attention is paid to evaluating the
uncertainties in derived radii -- a critical factor in assessing the
changes in stellar periods expected as PMS stars contract, absent external
regulation.

We find that PMS stars spanning an age range from $\sim$0.1 to 3 Myr do
not appear to conserve stellar angular momentum as they evolve down their
convective tracks, but instead preserve the same range of periods even
though they have contracted by nearly a factor of three. Our conclusions
apply both to stars which appear to lack the near-infrared excesses
diagnostic of circumstellar accretion disks and those that  show
conclusive evidence of disks.

Independent supporting evidence for this result follows both from analysis
of the relationship between observed periods and derived stellar radii
using other recently published surveys, and from recently published
projected rotational velocities observed for a sample of stars drawn from
the Orion Nebula Cluster itself (Rhode \etal\ 2001; RHM). These latter data
reveal a {\it decrease} of average \vsini\ with decreasing radius,
consistent with the results derived from the ONC Flanking Fields and NGC
2264 period studies.

Together, these results seem to require a mechanism that constrains young
stars to a constant range of periods from the time they first appear on
the stellar birthline to ages at least as great as $\sim$3 Myr.  We
discuss several candidate mechanisms, the most plausible of which appears
to be disk-locking.

\section{Observations}

This paper makes use of data sets for three young clusters:  the Orion
Nebula Cluster (ONC), the Flanking Fields (FF) that surround the ONC (cf.\
Rebull 2001), and NGC 2264.  Together, these clusters allow us to examine
changes in stellar angular momenta for stars with and without IR
signatures of disks having ages, masses, and radii spanning the ranges
$\sim 0.1-3$ Myr, $\sim 0.2-2$ \msun, and $\sim 1-5$ $R_{\sun}$
respectively.  In order to place stars in observed and reddening-corrected
color-magnitude (CMD) or Hertzsprung-Russell (HRD) diagrams, determine
stellar radii, and assess whether or not a star has a disk, we require
that spectral types along with $VI_{\rm C}K_s$ photometry be available for
all the stars included in the present study.

We use \lbol\ and \teff\ to calculate stellar $R$: \begin{equation} R^2 =
\frac{L_{\rm bol}}{4 \pi \sigma T_{\rm eff}^4} \end{equation} To obtain
\lbol\ and \teff\ from the observed \ic\ and \vi, we first deredden the
observed \vi\ using the observed spectral type to estimate intrinsic
colors.  The dereddened \ic$_0$ is calculated from the extinction at \ic,
which is given by \ai=1.61\evi.  \teff\ follows from the spectral type. 
We convert \ic$_0$ to \lbol\ using the approach described by Hillenbrand
(1997; H97); distance moduli were taken from the literature as noted in
the sections that follow.

To ensure internal consistency, we recalculated previously published
extinction and color excesses for each star included in our discussion by
combining photometry and spectra reported in the literature with a common
set of assumed photospheric colors and reddening laws.

\subsection{The ONC}
\label{sect:onc}

The ONC is nearby (470 pc; Genzel \etal\ 1981), compact (size $\sim$1 pc),
and young (typical stellar ages of $\sim$1$-$3 Myr; see, e.g.\ H97).   It
has been studied recently by Rhode \etal\ (2001; RHM), who measured values
of the projected rotational velocity ($v \sin i$) for 155 stars (with
upper limits for another 83). Observed $V$, $I_{\rm C}$, and $K_s$ colors
and spectral types for these stars are available from H97.   There are 202
stars for which both the photometry and measurements of $v\sin i$ are
available; they appear in Figures~\ref{fig:cmdobs}$-$\ref{fig:hrd}.  These
figures show the observed and dereddened CMDs for the stars along with
their positions in the \lbol$-$\teff\ plane (Hertzsprung-Russell Diagram,
HRD). About 60\% of the stars in the $v \sin i$ sample have near-infrared
($I_{\rm C}-K_s$) excesses indicative of a circumstellar disk, e.g.\ a
de-reddened $I_{\rm C}-K_s$ color exceeding expected photospheric values
by 0.3 mag, the conservative criterion adopted by H97 (see also
Hillenbrand \etal\ 1998) to select the most probable disk candidates.

A significant advantage of using $v \sin i$ data for studies of PMS
rotational properties is that line widths can be measured for {\it every}
PMS star to a limit set by spectral resolution; for the RHM data this
limit is 11 km s$^{-1}$, which corresponds to P $>$ 6 days for a typical
PMS star. By contrast, searches for spot-modulated periods have biases
arising from: (1) the sampling cadence of photometric observations (which
can be determined if the observation times are known); (2) the greater
difficulty in determining periods for accreting PMS stars in which
stochastic, accretion-driven photometric variability can overwhelm
periodic spot-modulated signals; and (3) the possibility that some ranges
in stellar rotation rate may not produce the large, long-lived spots
required to establish periodicity.  Together, these biases could well
mitigate the significant advantage in principle of periods in establishing
stellar rotation absent the complication of projection effects ($\sin i$).

RHM establish a result essential to studies of stellar angular momenta
based on observations of spot-modulated periods: there is no statistically
significant difference between the $v\sin i$ distributions of PMS stars
with known periods ($P$) and those without.  Hence there is {\it no
statistical bias introduced by determining the distribution of rotational
properties for that subset of PMS for which period determinations are
possible}.

RHM calculate $R\sin i$ values by combining observed $P$ and \vsini.  They
also establish another result critical to our analysis: agreement in the
mean between these $R\sin i$ values and the $R$ values calculated from
\lbol\ and \teff.  This result supports our use of estimates of $R$
derived from the apparent location of an individual star in the HRD as an
independent variable that can be used to estimate the effects of
evolution-driven changes in angular momentum.

\subsection{The Flanking Fields}
\label{sect:ff}

\begin{figure*}[tbp]
\epsscale{0.75}
\plotone{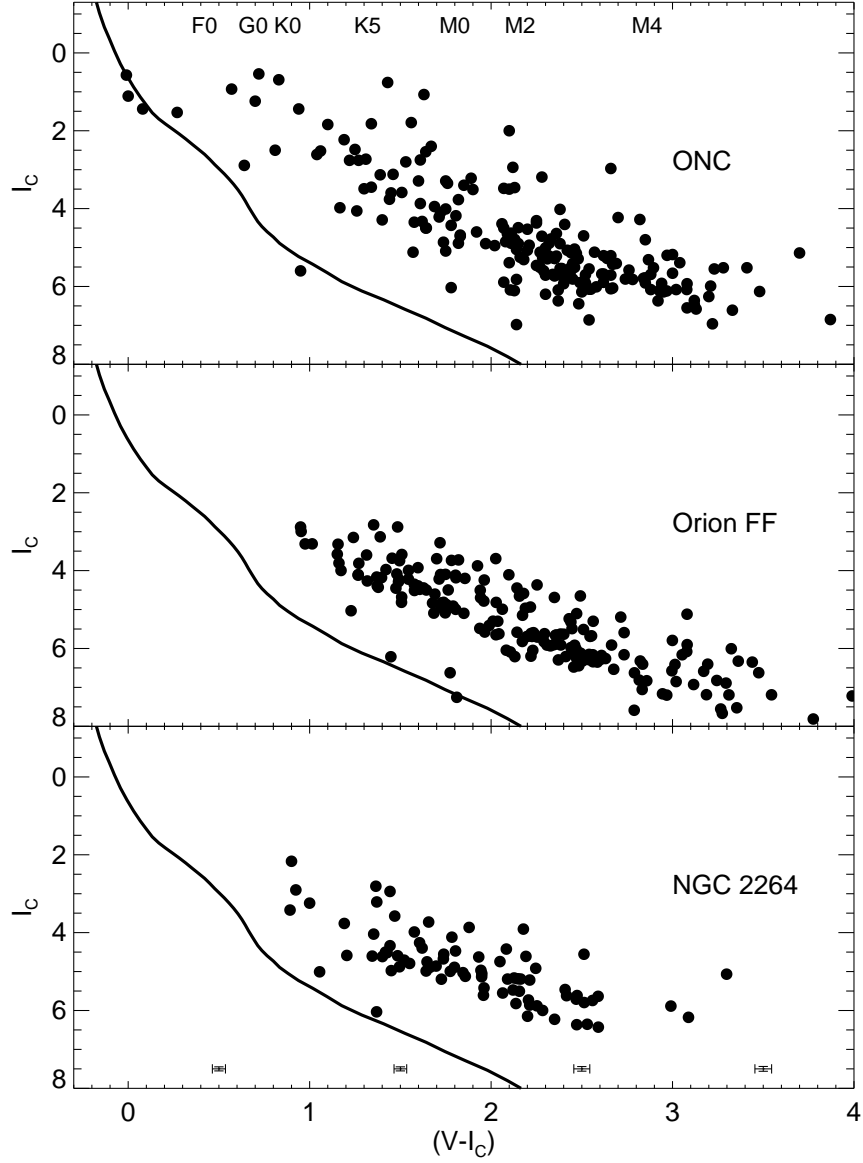}
\caption{Observed color-magnitude diagram (CMD) for each of the three
samples considered in this paper: the Orion Nebula Cluster itself, the
Orion Nebula Cluster Flanking Fields, and NGC 2264.  $I_{\rm C}$
magnitudes have been converted to absolute magnitudes (see text for
adopted distance moduli) for ease of comparison between the Orion samples
and NGC 2264.  ZAMS relation is heavy line, and \vi\ colors corresponding
to various spectral types are indicated in top panel.  In bottom panel,
typical standard error bars are indicated for points with \vi\ between 0$-$1,
1$-$2, 2$-$3, \& 3$-$4; see text. }
\label{fig:cmdobs}
\end{figure*}

\begin{figure*}[tbp]
\epsscale{0.75}
\plotone{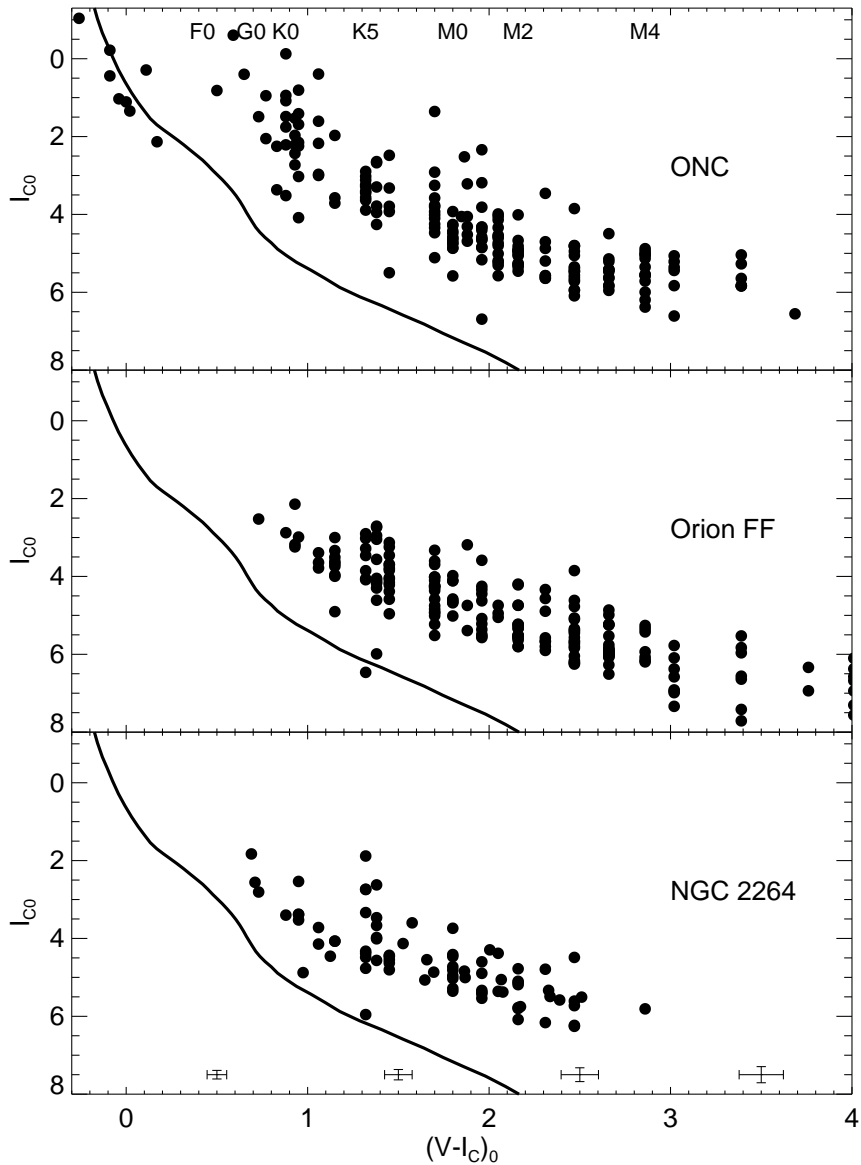}
\caption{Dereddened CMD for each of the three samples considered in this
paper: the Orion Nebula Cluster Flanking Fields, the Orion Nebula Cluster
itself, and NGC 2264.  $I_{\rm C}$ magnitudes have been corrected to
absolute magnitudes for ease of comparison between the Orion samples and
NGC 2264.  \vi\ values appear ``quantized'' because stars have been
reddening-corrected to the intrinsic \vi\ colors expected for their
spectral types. ZAMS relation is heavy line, and \vi\ colors of spectral
types are indicated in top panel.  In bottom panel, typical standard error
bars are indicated for points with \vi\ between 0$-$1, 1$-$2, 2$-$3, \&
3$-$4; see text. }
\label{fig:cmddered}
\end{figure*}

\begin{figure*}[tbp]
\epsscale{0.75}
\plotone{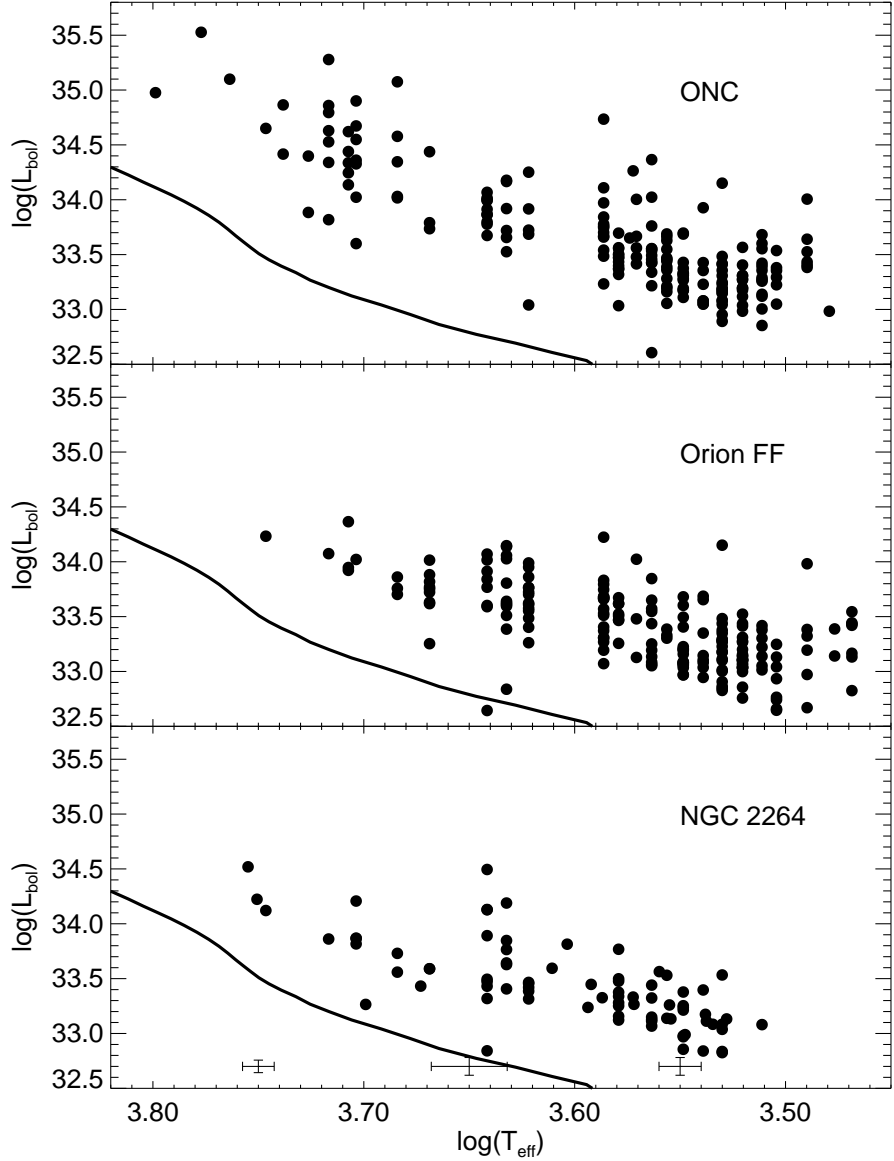}
\caption{Theoretical Hertzprung-Russell Diagrams (HRD) for each of the
three samples considered in this paper: the Orion Nebula Cluster Flanking
Fields, the Orion Nebula Cluster itself, and NGC 2264.  \lbol\ and \teff\
have been derived from dereddened colors following the formulas given in
Hillenbrand (1997).  ZAMS relation is heavy line, and \vi\ colors
corresponding to various spectral types are indicated on top panel.  In
bottom panel, typical standard error bars are indicated for points with
log \teff\ between 3.5$-$3.6, 3.6$-$3.7, \& 3.7$-$3.8; see text. }
\label{fig:hrd}
\end{figure*}

The Orion Flanking Fields (FF) were defined and studied by Rebull \etal\
(2000) and Rebull (2001; R01); we use $VI_{\rm C}K_s$ photometry, spectral
types, and $P$ from these papers, producing 197 stars which appear in
Figures~\ref{fig:cmdobs}$-$\ref{fig:hrd}.

Measures of stellar rotation for this sample are provided by
spot-modulated photometric periods ($P$).  Spot modulation is only found
in late-type stars, so unlike $v \sin i$ measurements,  which can be made
of any type star, $P$ observations are only obtainable for types mid-G
and later. Period determinations have uncertainties $\delta P <$1\%. The
$P$ measurements used here are complete over a broad range of periods, viz
$P\sim 25-0.3$d, corresponding to rotational velocities $v\sim 3-300$
km~s$^{-1}$, ranging from $\sim$1$-$100\% of the breakup velocity.

We have limited our primary set of $P$ data in Orion to the studies of the
FF cited above in order to ensure that we were working with a homogenous
data set that was complete over the largest possible range of periods. 
There are several other studies of the Orion region in the literature;
they are compared in Table~\ref{tab:otherstudies}.  Note that the ranges
of periods over which these studies are {\it complete} (not simply {\it
sensitive}) can vary widely.  Results from the other databases spanning
more restricted or different ranges of period nevertheless support the
conclusions reached from analysis of our primary database (R01), as will
be shown below.

Stars with observed periods in the FF are young and thought to be
associated with the ONC cluster for several reasons.  R01 used existing
membership studies to demonstrate that the possible field star
contamination in the sample of stars showing periodic variability is low,
hence supporting the assumption that stars in the FF with derived periods
are likely to be associated with the ONC. Further arguments that reinforce
this contention include that: (1) the CMD of stars in these fields
strongly resembles those from the ONC (see Rebull \etal\ 2000 and
Figures~\ref{fig:cmdobs}$-$\ref{fig:hrd}), and (2) the spatial
distribution of the periodic candidates is elongated N-S, along the
direction defined by the distribution of dense molecular material which
lies behind both the ONC and the FF. We thus adopt the same distance, 470
pc, for the FF stars as we did for the ONC sample.

R01 examines the $I_{\rm C}-K_s$ excesses for the FF sample as part of her
study of disk frequency and properties.  She finds that 30\% of the stars
in the FF sample have $I_{\rm C}-K_s$ excesses of 0.3 mag (or greater) 
above expected photospheric values -- the conservative criterion adopted
by H97 and Hillenbrand \etal\ (1998) as indicative of stars with high  
likelihood of a circumstellar accretion disk.

\subsection{NGC 2264}
\label{sect:2264}

Data for NGC 2264 are similar to those used for the FF sample.  Colors and
spectral types come from Rebull \etal\ (2002), and periods from Makidon
\etal\ (2002).  A total of 81 stars have spectral types, colors, and
spot-modulated periods.  These stars, appearing in
Figures~\ref{fig:cmdobs}$-$\ref{fig:hrd}, come from our study of NGC 2264
alone; the stars that we have in common with Kearns \& Herbst (1997, 1998)
are retrieved with identical periods.  About 20\% of the stars in this
sample have $I_{\rm C}-K_s$ excesses indicative of a circumstellar disk
using H97's conservative criterion, $I_{\rm C}-K_s$ excess $>$ 0.3 mag.

NGC 2264 (part of the Mon OB 1 association) is about twice as far away as
Orion; we used a distance modulus of 9.40, or 760 pc (Sung \etal\ 1997).
As in Orion, a molecular cloud located behind the cluster aids in blocking
background field stars and limiting the cluster depth to a small fraction
of the distance to the cluster.  Past studies have argued that  NGC 2264
is slightly older than the ONC at 3$-$5 Myr (\eg\ Sung \etal\ 1997).  Our
data suggest that NGC 2264 indeed  appears to contain fewer very young
stars when compared with the ONC, but Figure~\ref{fig:cmddered} also
suggests that the {\it range} of ages found in NGC 2264 is not dramatically
different from the ONC (Rebull \etal\ 2002). The value of the NGC 2264
sample for this study thus lies in its containing a larger {\it fraction}
of stars with ages comparable to 3 Myr than either the ONC or FF.

\section{Sources of error in $R$}
\label{sect:errors}

Key to our study is the ability to estimate stellar $R$ accurately enough
to track the changes in $P$ expected as PMS stars contract and evolve down
their convective tracks.  The procedure for calculating $R$ is described
in the previous section.  Sources of error in this calculation include
photometric uncertainty, the amplitude of stellar variability,
uncertainties in the reddening correction, errors in classification,
uncertainties in the intrinsic photospheric colors, the effects of
accretion, errors in the distance, and the presence of companions.  All of
these effects contribute to errors in \lbol\ and \teff\ and hence to
uncertainty in $R$.  In the remainder of this section, we provide a
thorough discussion of uncertainties involved in deriving \lbol, \teff,
and $R$ from the observed colors and spectral types.  Photometric and
spectral type errors are summarized in Table~\ref{tab:errors}.  Unless
otherwise specified, \lbol\ is in ergs cm$^{-2}$ s$^{-1}$, \teff\ in K,
and $R$ in $R_{\sun}$.

\subsection{Preliminaries}

Since \begin{equation} 2 \log R  = \log L_{\rm bol} - \log (4\pi\sigma) -
4\log T_{\rm eff} \end{equation} and for \begin{equation} x=au\pm bv
\end{equation} the propagation of errors is described by (Bevington \&
Robinson 1992) \begin{equation} \sigma_x^2 = a^2\sigma_u^2+b^2\sigma_v^2
\pm ab \sigma_{uv}^2, \end{equation} it follows that: \begin{equation}
(\delta \log R)^2 = \frac{1}{4}(\delta \log L_{\rm bol})^2 + 4 (\delta
\log T_{\rm eff})^2 \label{eqn:err} \end{equation} where we assume that
the errors are independent.  This assumption gives an upper bound to the
total error, since errors in spectral type produce partially offsetting
errors in \teff\ and \lbol.

\subsection{Errors in Brightness and Color}
\label{sect:photerr}

Photometric errors result from photon counting statistics, uncertainties
in atmospheric extinction corrections, and uncertainties in transforming
to a standard system.  These errors are smallest for the brightest stars
and largest for the faintest stars, but are typically $\sim$0.03 mag in
both \ic\ and \vi\ for most stars in our sample.

Photometric variability on scales ranging from millimagnitudes to
magnitudes is characteristic of young stars (e.g.\ Herbig 1954).  The
light curves of the samples with periods (the FF and NGC 2264 stars) have
typical amplitudes (mean-to-peak) of 0.03 mag in \ic\ for all but the
stars with spectral types M3 and later, which have somewhat larger
amplitudes. The \vsini\ sample was specifically selected to include
aperiodic as well as periodic variables.  We do not have monitoring
information on these specific stars, but the distribution of variations of
similar aperiodic stars in the FF and NGC 2264 is strongly peaked near
$\sim$0.03 mag as well.  We therefore take $\sim$0.03 mag as the typical
amplitude for all stars in our samples.

In our photometric studies of the FF and NGC 2264, we monitored in one
filter, \ic, only, and thus we do not have direct information on how \vi\
or $V$ changes with these periodic modulations.  Carpenter \etal\ (2001)
monitored stars including those in 2 of the 4 FF in $JHK_s$.  They find
that most (between 57 and 77\%) of their variable stars can be accounted
for by cool spot models, e.g. low-amplitude, nearly colorless
modulations.  Based on this, we assume that the amplitude of variation we
found at \ic\ also occurs in $V$, and that it does not affect \vi\ at
significant levels.

If we assume that \vi\ is affected only by uncertainties resulting from
photometric uncertainties, for stars with types between K5 and M2, we
derive uncertainties in luminosity  $\delta \log L_{\rm bol}\sim$0.04;
similarly $\delta \log T_{\rm eff} \sim 0.002$.  Using equation
\ref{eqn:err} above, $\delta \log R \sim 0.02$.

If we assume that \ic\ is affected by photometric uncertainties and by
variations with an amplitude of 0.03 mag, we similarly derive, for stars
with types between K5 and M2, $\delta \log L_{\rm bol} \sim 0.02$, and
$\delta \log T_{\rm eff} \sim 0$ (because changing \ic\ alone does not
change \teff), and thus $\delta \log R \sim 0.01$.  These uncertainties
increase with \vi\ primarily because the redder stars are also fainter.

\subsection{Reddening corrections, photospheric colors, and spectral
type uncertainties}
\label{sect:sptyerr}

We use spectral types to determine the intrinsic stellar colors and hence
to derive (interstellar) reddening.  The dominant error in this process is
the uncertainty in spectral classification.  Errors in typing have been
estimated (H97, Rebull \etal\ 2000) to be one subclass for the earlier
types (F, G, K) and at half a subclass for the later types (M).  For stars
that are still accreting material from their disks, the effects of veiling
on the observed photospheric spectrum result in classifications that are
systematically too early (e.g.\ H97).

Another potential source of error is uncertainty in the intrinsic
photospheric  colors.  It is standard practice to adopt ZAMS colors as the
best estimate for the true photospheric colors for these young,
low-gravity stars.  Our array of ZAMS colors includes measurements from
Bessell (1991), Leggett (1992), and Leggett \etal\ (1998).  This issue was
discussed in detail by Rebull \etal\ (2000), so we only summarize some of
the most important issues here.  (1) Colors for types $>$M4 are the most
uncertain.  (2) Chromospheric activity at levels typical of PMS stars  can
in principle affect PMS colors, but hot active regions are not likely to
significantly affect colors for passbands $V$ and redder (cf.\ Rebull
\etal\ 2000). (3) Large dark spots can also in principle affect both
colors and spectral types; their effects have not yet been quantified. (4)
Finally, the use of dwarf colors for these young subgiant stars may be
questionable.  To place an upper bound on this latter effect, we have
compared colors for dwarf stars with evolved stars of luminosity class I
and II.  Based on this comparison, we note that PMS stars are likely to be
slightly redder than dwarfs to about M1, and bluer for stars later than
that.  However, the size of this error is small compared to the color
change effected by assuming one spectral (sub-)class uncertainty in each
direction.  For example, a worst-case scenario occurs at M2, where the
difference between dwarf and giant \vi\ is $\sim$0.25 mag.  By comparison,
the difference between \vi\ colors of M2 and M2.5 dwarfs is $\sim$0.5 mag.
Since the true intrinsic colors of the PMS stars are closer to dwarfs than
stars of class I and II (c.f.\ H97), this error will be in reality even
smaller.

We have estimated the total error from all of the sources considered in
this subsection by shifting the observed spectral type one subclass (or
half a subclass for the M stars) blue and red, and then comparing the
resulting \lbol, \teff, and $R$.  Typical standard errors for stars in the
type range K5$-$M2 are  $\delta \log L_{\rm bol} \sim 0.06$, $\delta \log
T_{\rm eff} \sim 0.01$, and $\delta \log R \sim 0.04$. The uncertainties
in \teff\ increase as a function of type to mid-K owing to the uncertainty
in assigning spectral types and the slope of the relationship between
spectral type and \teff, but decrease from late K through the M-types,
because the precision of our assigned spectral types improves for these
late type stars.  The uncertainties in \lbol\ increase monotonically as a
function of type.

We note that stars in the ONC may suffer from an additional effect: a
range in the ratio of total to selective extinction that apparently
derives from a spatially variable mix of interstellar grain sizes in
the vicinity of the Trapezium.  For a typical \vi\ color excess, the
uncertainty driven by a change of a factor of two in the total to
selective absorption ratio will vary by $\sim$0.5 mag, which
corresponds to an uncertainty of log \lbol\ of $\sim$0.2 dex or log
$R$ of $\sim$0.1 dex.

\subsection{Accretion}

Material passing through a circumstellar accretion disk is funneled toward
the star along magnetospheric columns, and eventually strikes the
photosphere at supersonic speeds, producing hot ``accretion spots" via
shock heating.  Evidence of these accretion spots is manifest as
ultraviolet emission, well in excess of photospheric values for stars
having large accretion rates.  In recent papers, we have exploited excess
UV emission as one of several measures used to evaluate whether a young
star is surrounded by an accretion disk (\eg\ Rebull \etal\ 2000, Rebull
\etal\ 2002).  For stars with high accretion rates, the accretion spot
emission can be so strong as to affect $V$ and \ic\ colors as well. 
However, stars with accretion spot emission strong enough to affect $V$
and \ic\ are unlikely to be detected as periodic variables owing to the
dominance of large, random, accretion-driven luminosity changes, as
compared with lower amplitude spot-driven variations.  Quantitative
evidence of the effects of accretion on period detectability follows from
analysis of the sample of pre-main sequence stars in the Orion FF.  Of the
stars in the present sample with excesses derived directly from photometry
and spectroscopy, only 30\% have detectable periods; by contrast, of the
stars likely to be cluster members which lack UV excesses, 50\% have
periods.  Consequently, we do not believe that the periodic subsample
considered in this paper is dominated by stars with accretion rates
sufficiently high to affect $V$ and \ic\ colors.

We cannot estimate the role of accretion for the ONC sample.  While the
majority (60\%) of these stars have disks as judged by \ik\ excesses
$>$0.3 mag, we have no estimate of the distribution of UV excesses among
the sample, primarily because the high and variable background in this
region of the nebula precludes accurate ultraviolet photometry from the
ground.

In this latter context, we note the recent estimates of the effects of
accretion-driven changes on PMS star colors and derived luminosities
carried out by Hartmann (2001).  From analysis of a large sample of PMS
stars in the Taurus-Auriga region, he finds that  uncertainties in derived
luminosities owing to the effects of accretion are typically 0.06 dex in
log \lbol, or 0.03 dex in log $R$.

\subsection{Distance to and width of the clusters}

Uncertainty in distance to the cluster can result in a {\it systematic}
uncertainty in derived \lbol\ and $R$ values.  From the literature, the
distance to Orion is 470$\pm$70 pc; that uncertainty of 70 pc translates
directly to a {\it systematic} uncertainty of 0.07 dex in log $R$.
Similarly, NGC 2264 is at 760$\pm$30 pc, which translates to a {\it
systematic} uncertainty $\delta \log R$=0.03.  Note, however, that an
error in the average distance to a cluster simply introduces a systematic
{\it shift} in \lbol\ or $R$ for all the stars in the cluster.  An
uncertainty in distance does not therefore affect the {\it relative}
values of $R$ within the cluster, although it will affect comparisons of
clusters with each other.

The depths along the line of sight of the three groups of stars studied
here appear to be small compared to their distances. The apparent angular
size of the ONC is (generously) 0.5 deg, corresponding to 1\% of the
distance to the cluster.  The scatter in luminosity introduced by
variations in the distance to individual stars within the ONC is thus much
smaller than other sources of error; a 1\% error in \lbol\ corresponds to
less than 1\% error in $R$.  As argued above, in the case of the FF, the
presence of the molecular cloud and the evident association with and
similarity to the young stars in the central ONC all suggest that the
distance and depth of the cluster in these fields cannot be significantly
different from the distance and depth of the central ONC.  The cluster NGC
2264 also has a molecular cloud behind it and its depth along the line of
sight is again a small fraction of the distance to the cluster.

\subsection{Binaries}

The presence of binaries could significantly affect the measured \ic\ and
therefore the derived \lbol.  The worst-case scenario is two equal-mass
binaries, where the measured \lbol\ would be actually twice what it should
be for a single star; in this case, it would create a 0.3 dex in log
\lbol\ and thus a 0.15 dex change in log $R$.  However, the stars studied
here are unlikely to have companions with primary to secondary luminosity
ratios less than about 5:1.  For the ONC sample, RHM obtained spectra for
all of their stars and would have seen double lines either directly or in
the cross-correlation peaks had the primary:secondary luminosity ratios
been smaller than 5:1; they found only 7 binaries in their full
sample.  For the Orion FF and NGC 2264, the period sample is also unlikely
to be contaminated by unresolved binaries with similar companion
luminosities. Analysis of such putative binaries would reveal {\em two}
peaks in the power spectrum analysis of their light curves were the
primary:secondary luminosity ratio smaller than $\sim$5:1. Therefore we
estimate that the increase in log \lbol\ as a result of unrecognized
binaries is unlikely to exceed 0.10 for any of our samples, corresponding
to $\delta \log R \sim 0.05$.

\subsection{Final error estimates}

The errors in log R estimated above for each readily quantifiable source
are summarized in Table~\ref{tab:errors} for three different mass ranges. 
In addition, we estimate contributions from accretion (0.03 dex),
anomalous reddening appropriate to the Trapezium region (0.1 dex), and
binarity (0.05 dex).  For a typical star in our sample (spectral type
between K5 and M2), the combination of standard errors from photometry,
colors, spectra and reddening amount to 0.05 dex (c.f.
Table~\ref{tab:errors}). Hence, a reasonable {\it upper limit} to the
uncertainty in log $R$ for stars in this spectral type range would be
$\sqrt{0.05^2 + 0.03^2 + 0.1^2 + 0.05^2}=0.13$.

\subsection{Comparison to other error estimates}

These error estimates were derived independently of those discussed in
detail by Hartigan, Strom, \& Strom (1994); our error estimates of $\delta
\log L_{\rm bol}$ are identical at 0.08 dex, and our uncertainties in 
photometry and spectral types of $\delta \log T_{\rm eff}$ are essentially
identical (0.015 vs.\ 0.018 dex).

Hartmann (2001) has also provided a careful assessment of the
uncertainties in \lbol\ and \teff\ for stars in Taurus.  His final error
estimate for delta log \lbol\ ranged from 0.09$-$0.16 dex.  However, as
noted previously, he adopted an error of $\sim$0.06 dex in log \lbol\ as
typical of the uncertainties introduced by the effects of accretion on the
colors of classical T Tauri studies included in his study. Furthermore,
depth for the Taurus region compared to its distance is also much larger
than for our clusters, adding an additional uncertainty of 0.06 dex in log
\lbol\ for Hartmann's sample.  Even if we were to adopt Hartmann's
uncertainties (clearly an upper limit to that appropriate for our sample)
we would find that the error in log $R$ is less than 0.10 dex.

\subsection{External constraints on errors}
\label{sect:extconst}

As noted in \S\ref{sect:onc}, RHM estimate values of stellar radius from
(1) log \lbol\ and \teff\ ($R$(HRD)), and (2) from observed periods and
projected rotational velocities, \vsini\ ($R\sin i$). From 73 stars with
both measured $P$ and \vsini\ values, we deduce \begin{equation} \log
R\sin i = 0.93 (\pm 0.2) \log R({\rm HRD}) -0.19 (\pm 0.1) \end{equation}
The scatter in log $R$ for each star is $\sim$0.2 dex. Approximately 0.15
dex of this scatter arises naturally from the spread in $R\sin i$ values
introduced by projection effects. The remaining error source must amount
to $\sim$0.13 dex, a value consistent with our upper limit error estimate
for stars in the Trapezium region (see section 3.7 above). 

Another external check on our error estimates is provided by the analysis
of Hartigan \etal\ (1994).  These authors estimate luminosities for
individual members of PMS binary systems in the Taurus-Auriga region using
methodology similar to that described above. They then compare the
observed ratio $L$(primary) / $L$(secondary) with that expected if both
stars were born at the same time, and thus fall on the same (computed)
isochrone. For a sample of 29 stars, two thirds have luminosity ratios
within 0.25 dex of the expected value, while 90\% of the sample lies
within 0.4 dex.  Hence, subject to the assumption that the computed
isochrones are correct, this comparison suggests that {\it typical}
standard uncertainties in log $R$ should be 0.12 dex or less.

\section{Analysis}

If stars conserve (stellar) angular momentum ($J$) as they evolve down
their convective tracks, then we would expect that $J=I\omega$ would
remain constant (where $I$ is moment of inertia and $\omega$ is angular
velocity).  This is equivalent to the requirement that $J=MvR$ be constant
if $J$ is conserved in the outermost observable layer of the star.
Although we expect fully convective stars to rotate as solid bodies,
theoretical models of PMS stars (\eg\ Swenson \etal\ 1994)  predict that
changes in surface rotation rate as a star evolves down the convective
track are the same whether the star rotates as a solid body or conserves
$J$ locally (\ie\ $J$ conservation in spherical shells).  The reason is
that along the convective track contraction is nearly homologous, and
changes in $I$ directly track changes in $R^2$.  For a given mass, if $J$
is conserved, then $vR$ is constant and $P = 2\pi R/v \propto R^2$.  


In the following sections, we examine the relationships between $R$ and
$P$ and between $R$ and \vsini\ and compare these empirical relationships
with the simple prediction that stellar angular momentum is conserved, as
expected absent angular momentum loss.

\subsection{The FF and NGC 2264}

In Figures \ref{fig:pvsrff} and \ref{fig:pvsr2264}, we plot $P$ vs.\ $R$
for the stars in the FF and NGC 2264.  The data are shown for three
different spectral type groups, which has the advantage of effectively
segregating the samples by mass.  In this way, we can also show the data
for stars with spectral types M3 and later separately, since the errors in
$R$ are largest in this type range.  This dividing line is also
convenient, as (using D'Antona \& Mazzitelli models) it corresponds to 
$\sim$0.25$-$0.3 \msun; this is the location of the break in the
distribution of periods as found by HRHC and HBJM.

Reference to Figures \ref{fig:pvsrff} and \ref{fig:pvsr2264} (see also Fig
\ref{fig:pvsronc}) shows that the change in log $R$ is, depending on the
cluser, at least 0.6 and could be as large as 0.9.  Allowing for the
standard error of 0.13, it is therefore reasonable to assume that there
are real changes in log $R$ of {\it at least} a factor of 3, a result that
is consistent with the evidence for contraction obtained by RHM.  Figures
\ref{fig:pvsrff} and \ref{fig:pvsr2264} show that for all three mass
ranges in these two clusters, there is {\it no evidence} for the spin-up
that would be expected with decreasing $R$ and increasing age if stellar
angular momentum were conserved. Rather, they suggest {\it evolution at
constant $P$, or constant angular velocity}.

While Figures \ref{fig:pvsrff} and \ref{fig:pvsr2264} and the fits show no
change of {\it average} period as PMS stars contract, it is more difficult
to ascertain via inspection whether the {\it distribution} of periods
changes significantly.  To search for such a change, we plot in
Figure~\ref{fig:quartile} two histograms depicting the frequency
distribution of $P$ for stars in the upper and lower quartiles of derived
stellar $R$ for our sample.  Values of $< \log P >$ and $< \log R >$ for
the upper and lower quartiles are presented in Table~\ref{tab:quartiles};
note that the average log $R$ for the upper and lower quartiles are
0.47$\pm$0.01 and 0.13$\pm$0.01 respectively, whereas the average log $P$
is 0.50$\pm$0.06 and 0.58$\pm$0.05. Were all stars free to spin up in
response to contraction, we would expect the shape of these histograms to
be the same, but the peak among the lower quartile stars to shift to log
$P$ = 0.50 $-$ 0.68 = $-$0.18.

If some of the stars were `locked' to fixed periods, for example by disks,
while others were free to spin up, we would expect the latter objects to
populate an extended `tail' towards short periods in the solid lines in
Figure~\ref{fig:quartile}.  Both visual inspection and quantitative
comparison of these distributions via a K-S test reveal no evidence that
there is a significant difference.  The fraction of stars with periods in
the lowest (shortest period) quartile of the distributions in
Figure~\ref{fig:quartile}a is 13/52 (25\%) for the lower quartile in
radius and 19/53 (35\%) for the upper quartile.  Within Poisson
statistics, these fractions are identical.

We conclude from Figure~\ref{fig:quartile} and the above analysis that
there is no significant difference in the frequency distribution of
periods among samples of stars with average log $R \sim$0.47 and 0.13
dex (the upper and lower quartile in radius).  Hence, there is no evidence
of spinup among the bulk of our sample.

The errors in $P$ are negligible, and so the only way to invalidate this
result is to hypothesize that the errors in $R$ are much larger than we
have estimated--so large, in fact, that we cannot use $R$ to sort stars by
age.  If there were a wide mixture of ages at each $R$ value, then the
spin-up expected for conservation of angular momentum in diskless stars
would be masked by observational scatter.  



Our conclusion that there is no apparent increase in P as stars contract
finds additional support from analysis of other datasets sensitive to
more limited ranges of $P$ and/or $R$.

Periods for stars in the ONC can be found in HRHC, SMMV, R01, CHS, and
HBJM.  We can perform a similar analysis of all of these data together for
stars just in the ONC, defined for purposes here as the region studied by
H97.  Figure~\ref{fig:pvsronc} is similar to Figures~\ref{fig:pvsrff} and
\ref{fig:pvsr2264} but for all 292 stars with available periods, $V$, \ic,
and spectral types.  Figure~\ref{fig:quartile} presents a similar
analysis of the distribution of log $P$ for the upper and lower quartiles
of log $R$ for this ONC sample; $< \log P >$ and $< \log R >$ are noted
in Table~\ref{tab:quartiles}. The fraction of stars with periods in the
lowest (shortest period) quartile of the distributions in
Figure~\ref{fig:quartile}a is 21/73 (29\%) for the lower quartile in
radius and 18/73 (25\%) for the upper quartile.  Within Poisson
statistics, these fractions are identical to each other and to the results
obtained with the FF data.

We next explore whether these results differ among samples which appear to
be surrounded by disks and those which appear to lack disks.  To enable as
robust a comparison as possible, we consider stars from the FF sample
surrounded by accretion disks to meet the conservative criterion \ik\
excess $>$ 0.3 mag established by H97; those which lack disks are assumed
to have \ik\ excess $<$ 0.1 mag.  We exclude stars with 0.1 $<$ \ik\
excess $<$ 0.3 mag from the sample on the grounds that establishing the
presence or absence of a disk is less certain in these cases. 
Figure~\ref{fig:newquart2} is similar to Figure~\ref{fig:quartile} but for
stars with \ik\ excesses $>$ 0.3 and $<$ 0.1 mag.  As before,
Table~\ref{tab:quartiles} summarizes  $<\log P >$ and $<\log R >$.  We
conclude that there is no significant difference in the distribution of
periods among young stars which appear very likely to be surrounded by
disks and those that lack disks {\it based on \ik\ excesses alone.}  This
result holds for both young stars with large radii and their presumed
descendents which have contracted by a factor of three.

We can fit lines to the log $P$ vs. log $R$ relationships depicted in
Figures \ref{fig:pvsrff} and \ref{fig:pvsr2264} (and \ref{fig:pvsronc})
and to a wide variety of subsets thereof.  Within the scatter, fits to all
of the subsets return a value consistent with zero slope.  To obtain the
largest range of $R$ possible, we combined all of the data from all of the
clusters and all of the surveys.  (The disadvantage to doing this, in
addition to the surveys' different sensivities and completeness, is that
Orion and NGC 2264 are at different distances, and if there are errors in
the distance, they may affect the comparison.)  We broke the entire data
set into two samples, with \ik\ excesses $>$0.3 and $<$0.1.  The slope of
a fit to these data is $0.2\pm0.2$ for the subsample with no excesses and
$-0.1\pm0.2$ for the subsample with excesses.  These results are
consistent with a slope of zero (no change of $P$ with $R$), and clearly
inconsistent with the slope of 2 expected were stellar angular momentum
conserved as stars contract.

\subsection{The ONC}

If stars indeed evolve at constant $P$ to ages $t\sim$ 3 Myr as Figures
\ref{fig:pvsrff} and \ref{fig:pvsr2264} suggest, then $v/R$ should be
constant.  In other words, a solid prediction based on the observations of
rotation periods in NGC 2264 and the FF is that older stars with smaller
radii should be rotating more slowly than younger stars with larger
radii.  If, on the other hand, angular momentum ($J$) is conserved during
evolution, then $vR$ is constant.  The typical change in $R$ over the
populated portion of the convective tracks is about a factor of 3 (cf.\
Figures \ref{fig:pvsrff} and \ref{fig:pvsr2264}).  If we take a typical
value of $v$ to be about 35 km s$^{-1}$ for the youngest stars (see
results of RHM below), then if $J$ is conserved, the oldest stars should
be rotating at more than 100 km s$^{-1}$.  If angular velocity is
conserved, then the oldest stars should be rotating at about 12 km
s$^{-1}$.

We can make use of the RHM dataset to search for evolution-driven trends
in rotational veclocity.  To do this, we have followed RHM and sorted the
stars by \lbol\ and \teff.  Within each (\teff,\lbol) `box', we average
the values of \vsini. This approach is necessary because there is
substantial scatter in the relationship between \vsini\ and $R$ among the
stars in this sample arising from (1) the wide range in rotational
velocity at fixed $R$ (examination of the range of observed periods at
fixed radius suggests an expected range of a factor of 10), and
(2) the range of inclination ($\sin i$) values.

In Figure \ref{fig:hrdgrid}, we superpose the (\teff,\lbol) `boxes' on an
HR diagram.  Also plotted are the stars observed by RHM and a set of
evolutionary tracks. This diagram shows that the stars observed in the
Rhode \etal\ ONC survey span the mass range $\sim$0.2$-$2.0 \msun.  Note
also that stars evolve down convective tracks at nearly constant \teff.
The RHM approach of sorting the stars by \lbol\ and \teff\  also
approximately sorts them by age and mass.  Each vertical column in the
grid shown in Figure \ref{fig:hrdgrid} spans a range of masses of about
30\% around the mean mass.  The convective tracks are not exactly vertical
in this diagram, and so some stars of a given mass may move from one
column to the next adjacent one as they evolve down their tracks. 
However, given that $<$\vsini$>$ at a given luminosity shows no systematic
trend with log \teff\ (see Figure 10), then the small change in average
mass with decreasing \lbol\ within a \teff\ column will not bias the results
significantly.   

As discussed above, the errors in log \lbol\ and log \teff\ are estimated
to be 0.08 and 0.015 respectively.  The width of the bins into which the
stars were placed are 0.37 in log \lbol\ and 0.048 in log \teff.  It is
therefore unlikely that the errors in deriving log \lbol\ and log \teff\
are large enough to move a star further than to a bin just adjacent to its
true location.  In most of the mass ranges studied, the stars span nearly
1.5 in log \lbol, and so the errors in log \lbol\ are not large enough to
mask systematic trends in $<$\vsini$>$ as a function of \lbol.  Because
(cf.\ Figure \ref{fig:hrdgrid2}) there are no systematic trends in \vsini\
with \teff\ at constant \lbol, errors in \teff\ are not large enough to
introduce or mask any systematic trends.

Figure \ref{fig:hrdgrid2} shows this same rectangular grid, but now with
lines of constant radius superposed.  Comparing this figure with Figure
\ref{fig:hrdgrid} shows that the data for stars in a given mass range
typically span a factor of $\sim$3 in radius.  The average value of
\vsini\ for the stars within each of the grid boxes is also shown.  The
data are summarized in Table \ref{tab:vsini}, which gives $<$\vsini$>$,
its standard deviation, and the number of stars in each grid box.

Examination of Figure \ref{fig:hrdgrid2} shows that stars near the tops of
the convective tracks have \vsini\ typically about 35 km s$^{-1}$.  Stars
near the bottom of the tracks have typical \vsini\ of about 20 km
s$^{-1}$--{\it far smaller than the 105 km s$^{-1}$ that would be expected
for conservation of stellar angular momentum.}  Instead, the observed
value of 20 km s$^{-1}$ is about 50\% higher than the 12 km s$^{-1}$ that
would be expected for conservation of angular velocity.  However, these
results are biased toward minimizing  the decrease in \vsini\ with $R$. 
In calculating these averages, we have used 11 km s$^{-1}$ for all stars
rotating at or below the resolution limit set by the spectra analyzed by
RHM.  Typically 20\% or fewer of the stars at the tops of the convective
tracks are rotating at or below this limit, and so $<$\vsini$>$ for the
youngest stars should not be significantly overestimated because of the
inclusion of stars with measured upper limits.  However, {\it more than
half} of the stars at the bottoms of the tracks rotate at or below 11 km
s$^{-1}$, and so the values of $<$\vsini$>$ quoted for the oldest stars
are too high.  This increase in the fraction of stars with \vsini $<$ 11
km s$^{-1}$ is another indicator that stellar angular momentum is lost as
stars evolve down their convective tracks.

Note, too, that the decrease in $<$\vsini$>$ with decreasing radius and
increasing age is found independently for each of the columns in Figure
\ref{fig:hrdgrid2}.

We can estimate the significance of this result by plotting the values of
$<$\vsini$>$ for each of the (\teff,\lbol) boxes against the
average value $<R>$ for each box; this relationship is shown in
Figure~\ref{fig:vsinir}. Note (c.f.\ Fig.~\ref{fig:hrdgrid2}) that
$<$\vsini$>$ is nearly constant along lines of constant $R$, and we can
therefore  combine the data for stars of all masses spanned by the
RHM sample.  From Fig.~\ref{fig:vsinir}, we derive a formal relationship
\begin{equation} \log <v\sin i> = 0.49 (\pm0.11) \log <R> + 1.15 (\pm0.05)
\end{equation} assuming uncertainties of 25\% in $<$\vsini$>$ and in log
$<R>$. Upper limits are set to 11 km s$^{-1}$ in computing $<$\vsini$>$
for each box.

The positive slope clearly shows that the rotation of stars in this sample
{\it slows down} as stellar radii {\it decrease}.  The slope of the
relation is clearly inconsistent with the slope of $-1$ that would be
expected were stars conserving angular momentum as they evolve.

Table \ref{tab:vrvr} gives for each grid box the product $<$\vsini$>R$,
which should remain constant down each column if angular momentum is
conserved, and $<$\vsini$>/R$, which should remain constant if angular
velocity is conserved.  Within a factor of 2 or better (especially given
the 11 km s$^{-1}$ upper limit on \vsini), angular velocity is conserved
for this ensemble of stars.  If we calculate rotational velocities for the
FF and 2264 samples using the derived values of $R$ and observed $P$, we
find the same trend toward decreasing rotation with decreasing R as in the
ONC.

Since 60\% of the ONC stars appear to have disks (based on having observed
\ik\ excesses $>$ 0.3 mag), one might conclude on the basis of this sample
alone that we have found strong evidence for disk-locking -- the most
plausible explanation for the apparent decrease in stellar angular
momentum with decreasing radius/increasing age.   As we have seen,
however, when we looked at the period data for the FF and NGC 2264, we
found similar evidence for loss of stellar angular momentum in samples of
PMS stars that appear to be dominated by objects that lack detectable
disks using the same detection criteria.

\subsection{Summary of the Observations}

PMS stars appear to preserve the same distribution of periods over radii
spanning 1$-$5 $R_{\sun}$, corresponding to ages 0.1-3 Myr.  The fact
that several different mass groupings of stars in all three clusters show
this same result, combined with the fact that the independent data sets
for the each cluster are characterized by different sources of error,
appear to make this a robust conclusion.

If correct, our results require that PMS stars lose angular momentum
over this age and radius range. We explore several candidate mechanisms
below.

\section{Discussion}

\subsection{Disk Locking}

Disk locking provides the most plausible explanation for the above
results. However, as noted above, we find that the bulk of our sample lack
robust evidence of disks, at least as measured by Hillenbrand's
conservative criterion, \ik\ excess $>$ 0.3 mag. Smaller values of \ik\
excess are expected if (1) the optically thick accretion disks that
produce \ik\ excesses are viewed nearly equator on; or (2) these disks
have large inner holes resulting from interruption of the accretion disk
by the stellar magnetosphere at radii exceeding 5 $R_*$ -- a possible
outcome if the stellar magnetic field is much larger than the few
kilogauss typical of TTS or if the disk accretion rate is much smaller
than the value of 10$^{-8}$ typical of young TTS.  In the former case, the
radiating area of the disk projected in the observer's direction is small,
thus resulting in small near-IR excesses, while in the latter case, the
inner disk lacks dust heated to temperatures sufficient to produce a \ik\
excess. Moreover, observational errors in estimating the \ik\ excess could
result in missing a significant fraction of stars with relatively small
excesses.

To explore the possibility that our sample might contain a significant
fraction of disks with small near-IR excesses, we recalculated the
fraction of disks for N2264, the ONC and the FF adopting as a criterion
\ik\ excess $>$ 0.1 mag. The computed fractions increased from 18, 35, and
58\% for \ik\ excess $>$ 0.3 to 59, 71, and 78\%. This suggests that it is
at least {\it plausible} that a much larger fraction of our sample could
be surrounded by disks, and hence that disk locking might well account for
the apparent decrease in stellar angular momenta as PMS stars contract.

Support for this speculation is also found from the recent study by Lada
\etal\ (2000). These authors undertook an $L$-band search for disks,
noting (1) the much larger excesses above photospheric levels expected at
$L$; and (2) the relative insensitivity of $L$ to the presence of small
inner disk holes. They found that about 45 percent of the M-type stars in
the Trapezium had $K$-band excesses as judged from their location in the
$J-H/H-K$ diagram but that about 80 percent of them had excesses at $L$.

A second search for disks with large inner holes was conducted by Stassun
\etal\ (2001), who searched for 10 $\mu$m excesses in stars in
Taurus-Auriga and Orion that lack \ik\ excesses.  In this case, only 3 of
32 stars showed evidence of 10 $\mu$m excesses, and on this basis the
authors argue that most of the stars without near-IR excesses do indeed
lack disks.

From these two studies, it is therefore difficult to estimate what fraction
of stars surrounded by disks are missed using \ik\ excess as a disk
discriminant.

Thus, while it appears tempting to argue that disk-locking accounts for
the apparent immutability of the period distribution among contracting PMS
stars, more work is needed in order to establish samples of PMS stars for
which periods are available and for which there is {\it definitive}
evidence establishing the presence or absence of accretion disks.

SIRTF observations promise the sensitivity for rapid and accurate 3$-$10
$\mu$m surveys capable of identifying disk candidates unambiguously. SIRTF
should also establish the (unlikely but still plausible) existence of
gaseous accretion disks in which all but a fraction of an Earth mass of
the small dust grains responsible for producing excess IR emission have
been agglomerated into larger bodies.

\subsection{Disk-Locking Combined with Birthline Effects}

Could the combination of similar period distribution for stars of
apparently different ages/radii and the lack of clear distinction in the
period distribution among stars with and without disks be explained
through some other mechanism?

As one possibility, suppose that (1) stars in Orion and NGC 2264 were
born in a single burst of star formation $\sim$1 Myr ago; (2) the range in
$L$ (equivalently $R$) for the PMS stars in these clusters does not
reflect evolution down convective tracks, but rather differences in mass
accretion rate (\mdot) that force stars to evolve along different
`birthlines' and therefore to arrive at different initial luminosities
along the convective tracks (Palla \& Stahler 1992); and (3) PMS stars are
locked to a particular $P$ so long as they are surrounded by accretion
disks. In this picture, the observed distribution of stars along a
convective track for a given mass then reflects a range of \mdot:
protostars with higher \mdot\ have larger initial radii and lie higher in
the color-magnitude diagram following the end of the accretion phase (see
also Hartmann \etal\ 1997). The similarity of $P$ at different $R$ then
follows from the assumption that stars are locked to their disks for much
(nearly all) of their $\sim$1 Myr accretion phase.

The main difficulty with this proposal lies with the additional
requirement that stellar $L$ cannot have decreased much since the stars
were deposited on their convective tracks.  If stars that formed through
high \mdot\ and were initially deposited high on their convective tracks
had subsequently evolved downwards, we would expect to see a `tail' of
stars that have spun up to shorter $P$ mixed in with the ensemble of
objects that started their evolution at smaller $R$.  We do not (see
Figure~\ref{fig:quartile}).  The only way to resolve this contradiction
and still maintain conservation of  stellar $J$ would be to identify a
mechanism for halting or slowing evolution of stars deposited high on
their convective tracks for a time comparable to at least $\sim$3 Myr.

\subsection{Angular Momentum Loss via Stellar Winds}

Stellar winds loaded onto open magnetic field lines can exert a spindown
torque on stars.  However, this mechanism appears to be ineffective for
PMS stars because, during this phase of evolution, the timescale for
spindown exceeds the evolutionary timescale by a few orders of magnitude
(\eg\ MacGregor and Charbonneau 1994) -- fully convective stars are
assumed to rotate as solid bodies, and the wind must slow down the entire
star.

Calculations (\eg\ Kawaler 1988) show that the rate of change of $J$
depends on the product of the mass loss rate and the square of the Alfven
radius, but the Alfven radius varies inversely as some power of the mass
loss rate, with the specific power depending on the configuration of the
magnetic field.  For a field geometry ``intermediate" between a dipolar
and a radial field, $dJ/dt$ does  not depend on mass loss rate at all
(Bouvier \etal\ 1997).  The only circumstance under which  magnetic winds
could play an important role in slowing the rotation of PMS stars would be
if there were some way in which to decouple the outer layers of the star
from the interior, \ie\ if $J$ were conserved locally, and the wind had to
slow down only a thin outer layer of the star.  Such decoupling is not
expected for fully convective stars.

A further challenge to any wind-driven $J$ loss mechanism is the
additional requirement that significant $J$ loss would have to cease on
timescales of no more than a few Myr in order to account for the
significant population of stars that ultimately arrive on the main
sequence as the rapid rotators observed in young clusters such as
$\alpha$Persei (\eg\ Stauffer \etal\ 1989); such stars require spinup from
the PMS to the ZAMS.

\subsection{Angular Momentum Loss via Tidally-Locked Planetary Companions}

Recent theoretical models indicate that tidal locking between a close-in
Jupiter-mass planet and the parent star can transfer spin $J$ from the
star to the orbital $J$ of the planet, thus slowing the rotation of the
star while driving the  planet into a larger orbit.

This effect is not large enough to account for the observations reported
here.  The difficulty lies in the apparent incompatibility of two
requirements: (1) the putative planet must be located close enough to the
star to produce significant tidal distortion; and (2) the planet must be
located far enough from the star to dominate the $J$ of the system (and
thus be able to create significant changes in stellar $J$ with only modest
orbital evolution).  A simple calculation (\cf\ Trilling \etal\ 1998) in
which we assume a Jupiter-mass planet orbiting a PMS star at a distance of
0.1 AU suggests that $J$ regulation by tidal locking to such a planet
fails by 2 orders of magnitude.

\section{Summary}

The observations of PMS stars discussed here span an age range of about 3
Myr and a change in stellar radius of about a factor of 3. Surprisingly,
observations of both rotation periods and projected rotational velocities
for a sample of several hundred stars show that most of the stars,
whether surrounded by observable disks or not, preserve constant angular
velocity $\omega$ to within a factor of 2 as they evolve downward along
their convective tracks -- and thus must lose stellar angular momentum $J$
in direct proportion to $R^2$.

A recent theoretical paper by Tinker, Pinsonneault, \& Terndrup (2002)
also concluded that stars must lose significant amounts of angular
momentum during the first few million years after they arrive at the
birthline.  These authors  assumed that the ONC stars, when they reach
an age of 120 Myr, should produce a rotational velocity distribution that
looks like the distribution in the Pleiades, and then calculated how the
angular momentum of the ONC stars would have to change with time to
achieve this outcome.  Tinker \etal\ concluded that there must be a
mechanism in addition to stellar winds that produces loss of stellar angular
momentum.  They further showed that, if this mechanism is disk-locking,
then it must be effective for {\it all} stars up to an age of about 3 Myr,
or alternatively disks must have a {\it range of lifetimes} over which
they serve as effective brakes, and that these lifetimes extend up to 6
Myr in some stars.

\begin{figure*}[tbp]
\epsscale{0.75}
\plotone{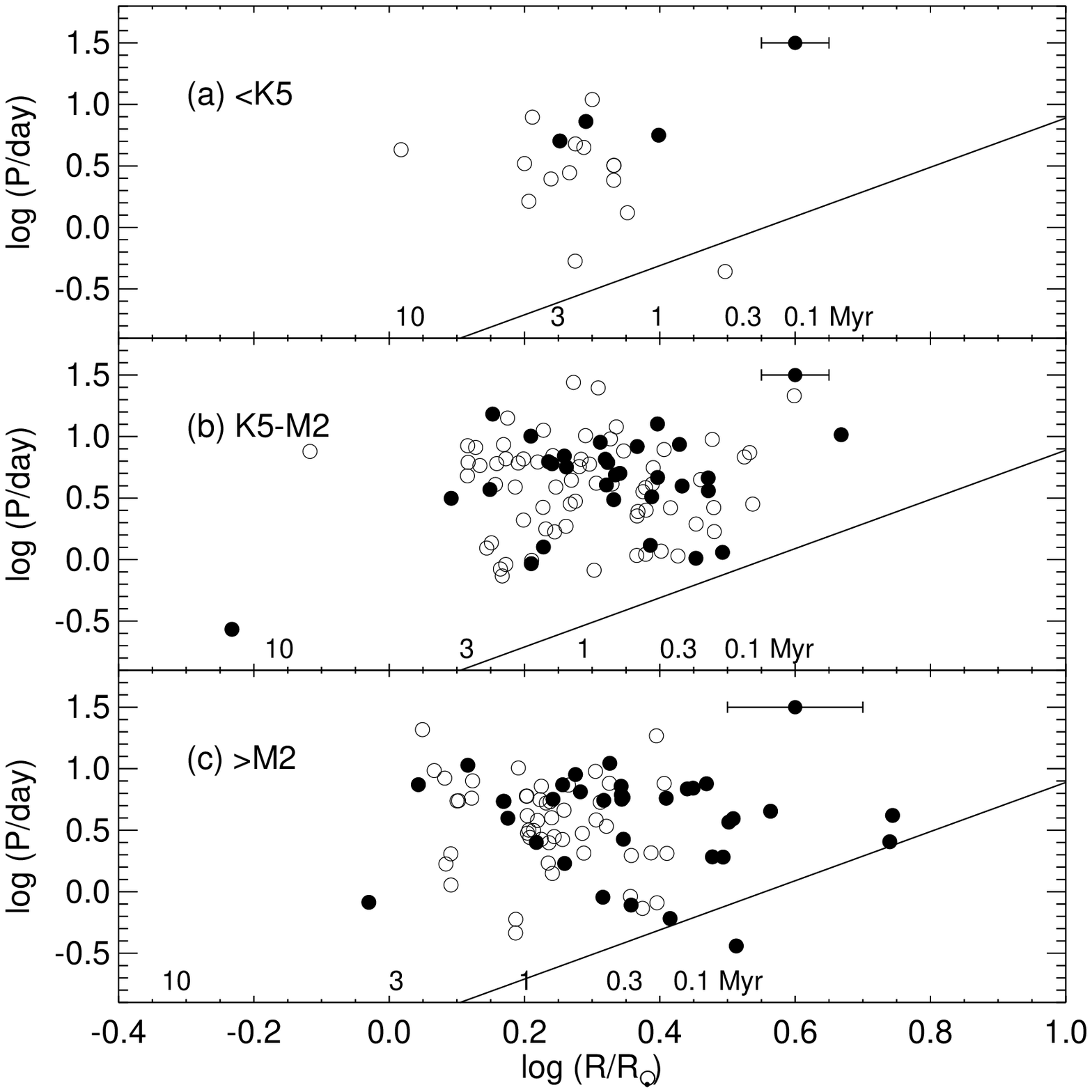}
\caption{Period vs.\ radius for stars in the FF.  (a) stars with types K5
and earlier; (b) stars with types K5-M2; (c) stars with types M3 and
later.  Open circles are stars without \ik\ excesses (i.e.\ without disks)
and solid circles are stars with excesses (i.e.\ with disks). Typical
standard error bars for each type range are indicated.  The line in each
panel indicates the slope expected if stellar angular momentum is
conserved. }
\label{fig:pvsrff}
\end{figure*}

\begin{figure*}[tbp]
\epsscale{0.75}
\plotone{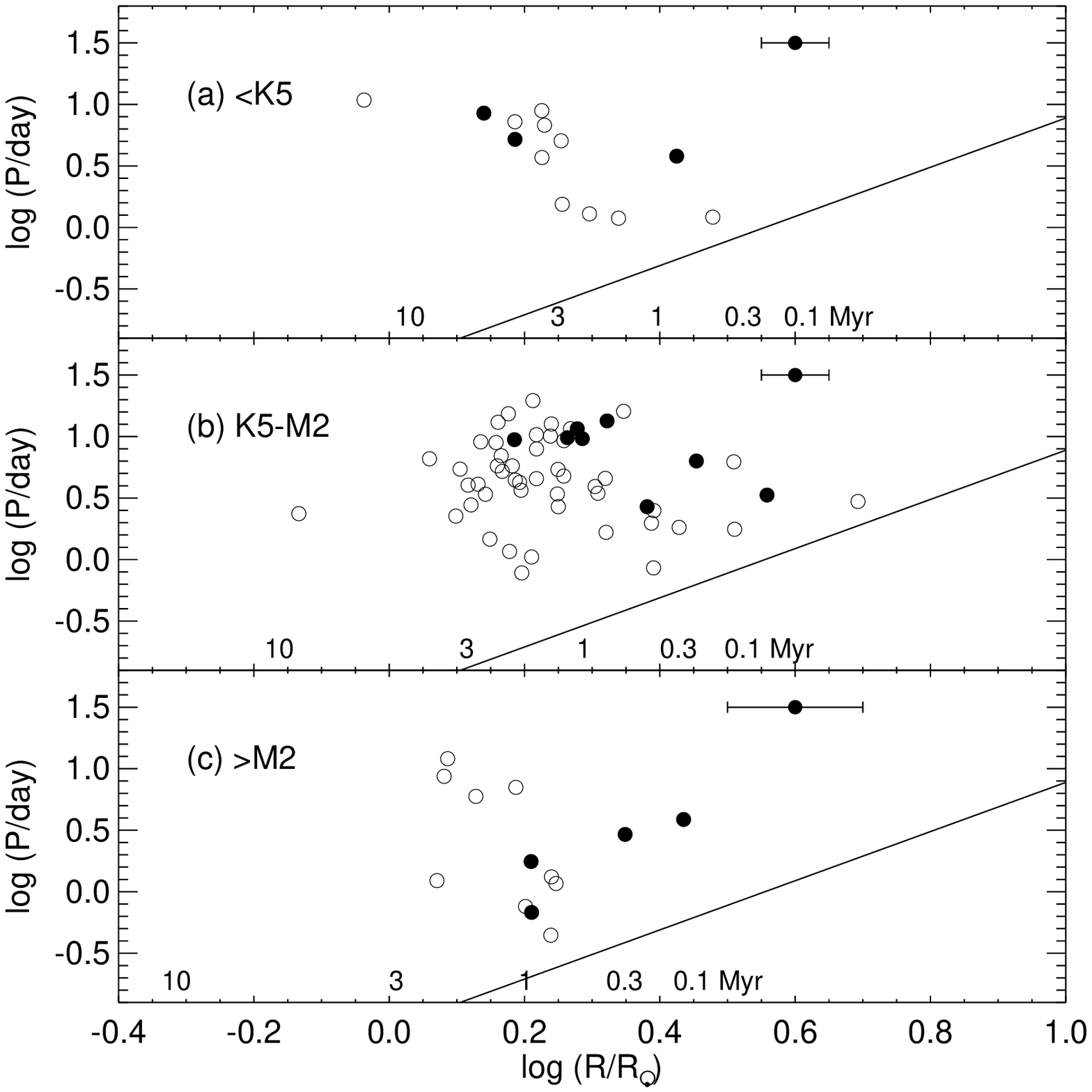}
\caption{Period vs.\ radius for stars in NGC 2264.  (a) stars with types
K5 and earlier; (b) stars with types K5-M2; (c) stars with types M3 and
later. Open circles are stars without \ik\ excesses (i.e.\ without disks)
and solid circles are stars with excesses (i.e.\ with disks).  Typical
standard error bars for each type range are indicated.  The line in each
panel indicates the slope expected if stellar angular momentum is
conserved. }
\label{fig:pvsr2264}
\end{figure*}

\begin{figure*}[tbp]
\epsscale{0.75}
\plotone{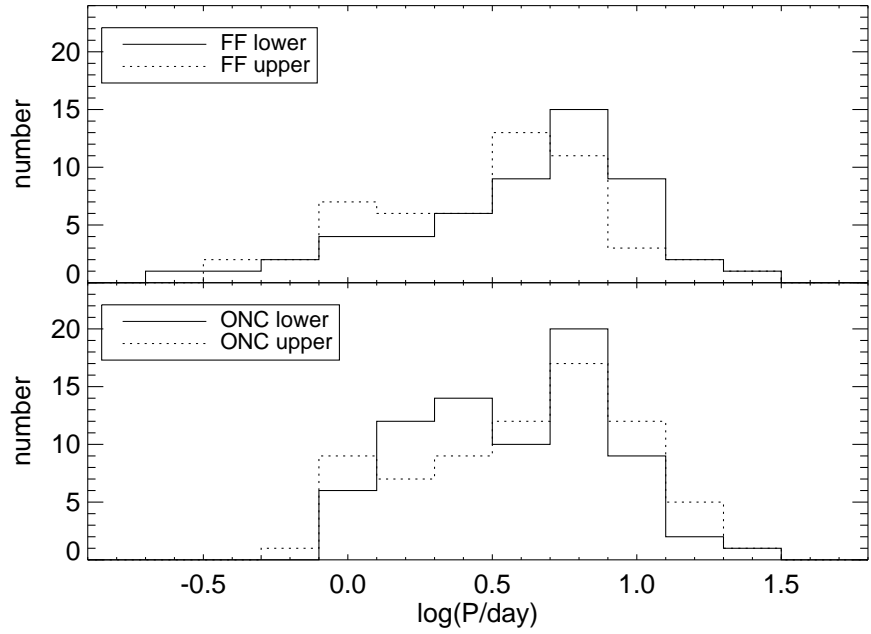}
\caption{Histograms of distributions of periods for stars in (a) the FF
and (b) the ONC for the upper and lower quartiles in radius.  There is no
difference in the period distributions.  See Table~\ref{tab:quartiles} for 
$<\log P >$, and $<\log R >$.  }
\label{fig:quartile}
\end{figure*}

\begin{figure*}[tbp]
\epsscale{0.75}
\plotone{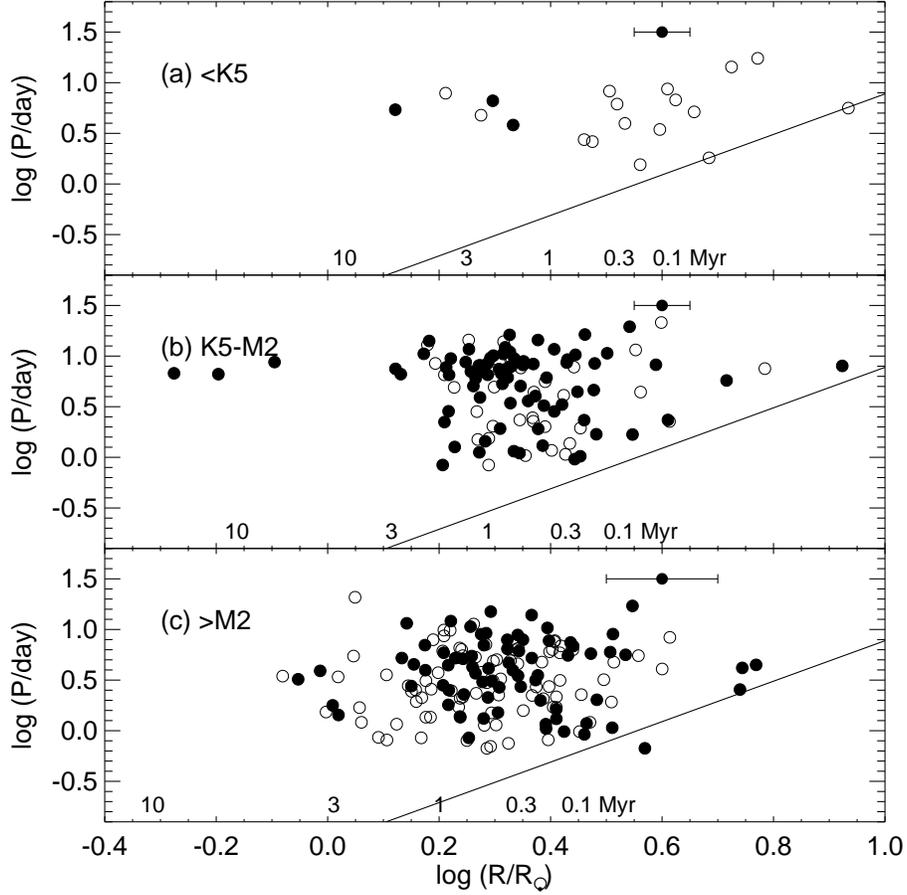}
\caption{Period vs.\ radius for stars from the ONC, defined as the region
covered by H97, periods taken from HRHC, SMMV, R01. CHS, and HBJM.  (a)
stars with types K5 and earlier; (b) stars with types K5-M2; (c) stars
with types M3 and later. Open circles are stars without \ik\ excesses
(i.e.\ without disks) and solid circles are stars with excesses (i.e.\
with disks).  Typical standard error bars for each type range are
indicated.  The line in each panel indicates the slope expected if stellar
angular momentum is conserved. }
\label{fig:pvsronc}
\end{figure*}

\begin{figure*}[tbp]
\epsscale{0.75}
\plotone{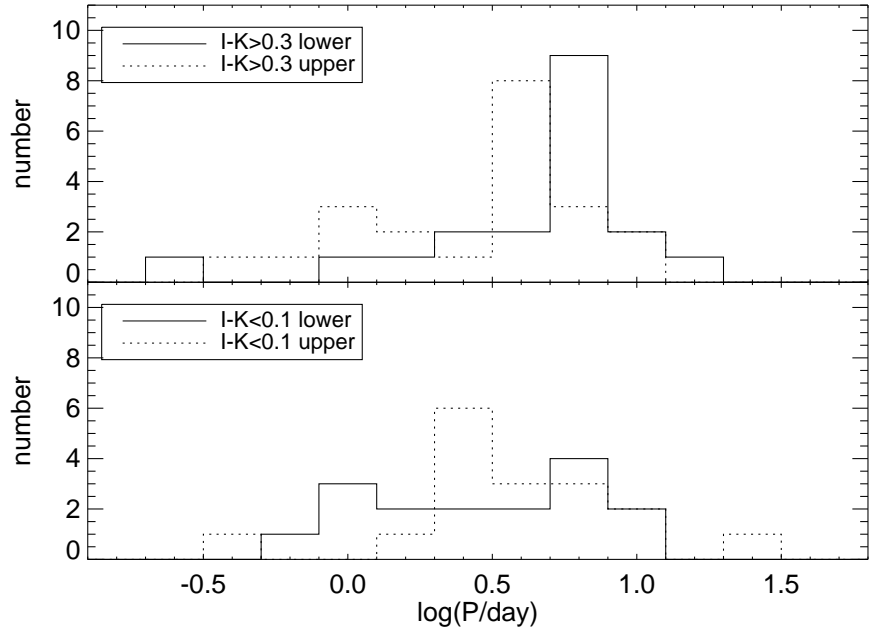}
\caption{Histograms of distributions of periods for stars in the FF with
(a) \ik\ excess $>$0.3 mag and (b) $<$0.1 mag.  There is no difference in
the period distributions.  See Table~\ref{tab:quartiles} for  $<\log P >$,
and $<\log R >$. }
\label{fig:newquart2}
\end{figure*}

\begin{figure*}[tbp]
\epsscale{0.75}
\plotone{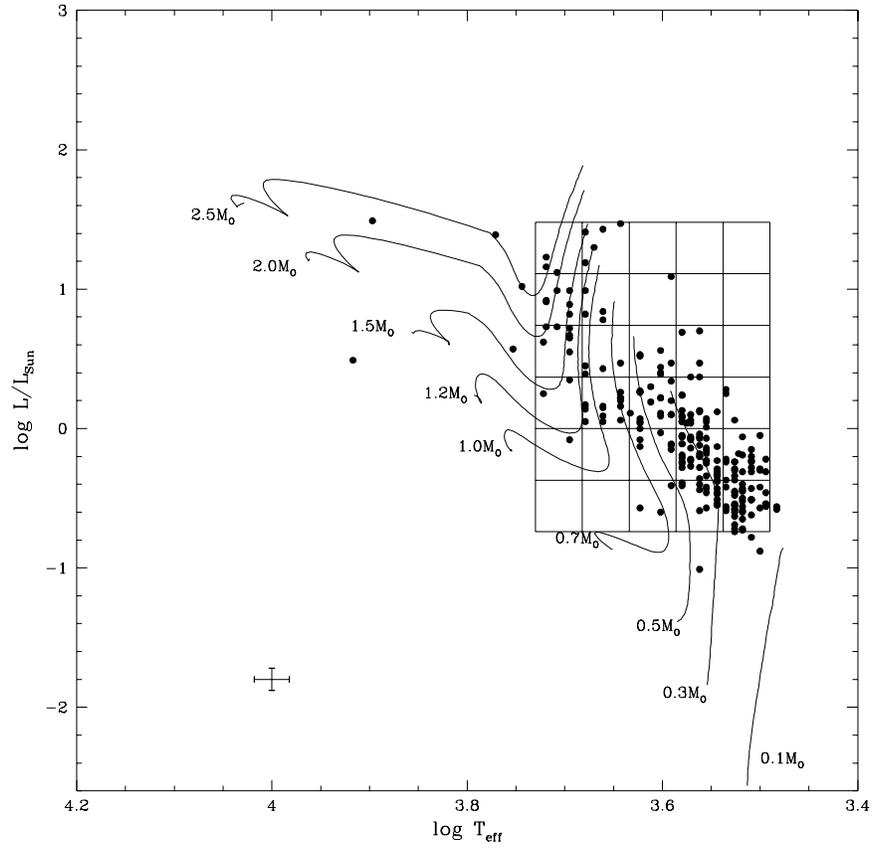}
\caption{The HRD for the ONC stars.  The evolutionary tracks are from
D'Antona and Mazzitelli (1994).  The grid shows the bins in the log \lbol\
and log \teff\ plane into which we sorted the stars in order to look for
trends in $<$\vsini$>$ with decreasing luminosity and increasing age.  The
estimated error bars for log \lbol\ and log \teff\ are given in the lower
left corner of the plot.}
\label{fig:hrdgrid}
\end{figure*}

\begin{figure*}[tbp]
\epsscale{0.75}
\plotone{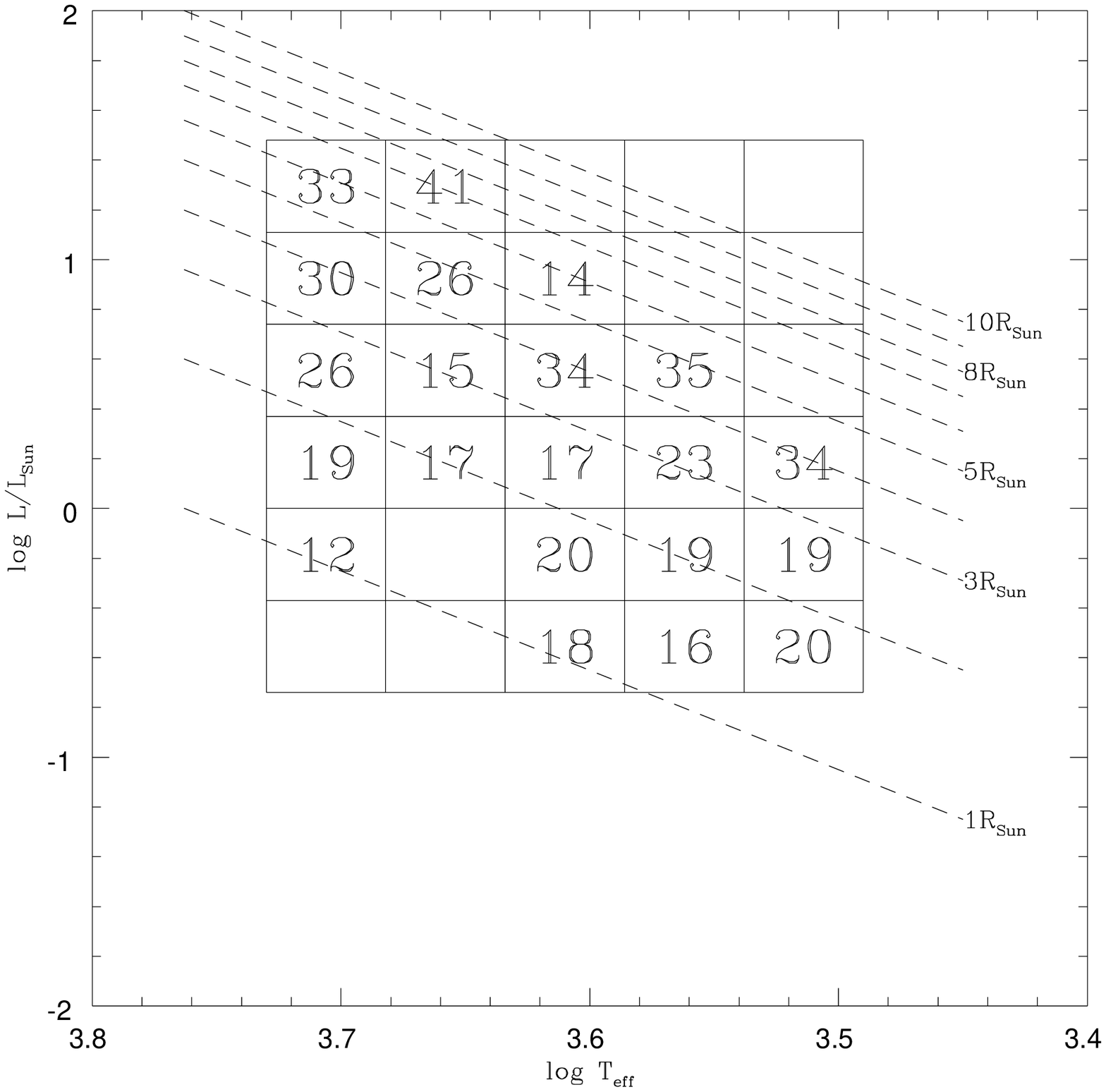}
\caption{The values of $<$\vsini$>$ in km s$^{-1}$ as a function of
position in the log \lbol\ and log \teff\ plane.  Lines of constant radius
are shown as dashed lines.  Note that the trend in all columns is toward
decreasing $<$\vsini$>$ with decreasing luminosity and decreasing radius,
which indicates that angular momentum must be lost as stars evolve down
their convective tracks.  Table \ref{tab:vsini} indicates how many stars
fall within each box in the grid and gives the statistical errors in the
values of $<$\vsini$>$.  Note that there is only one star in the top box
in the middle column.}
\label{fig:hrdgrid2}
\end{figure*}

\begin{figure*}[tbp]
\epsscale{0.75}
\plotone{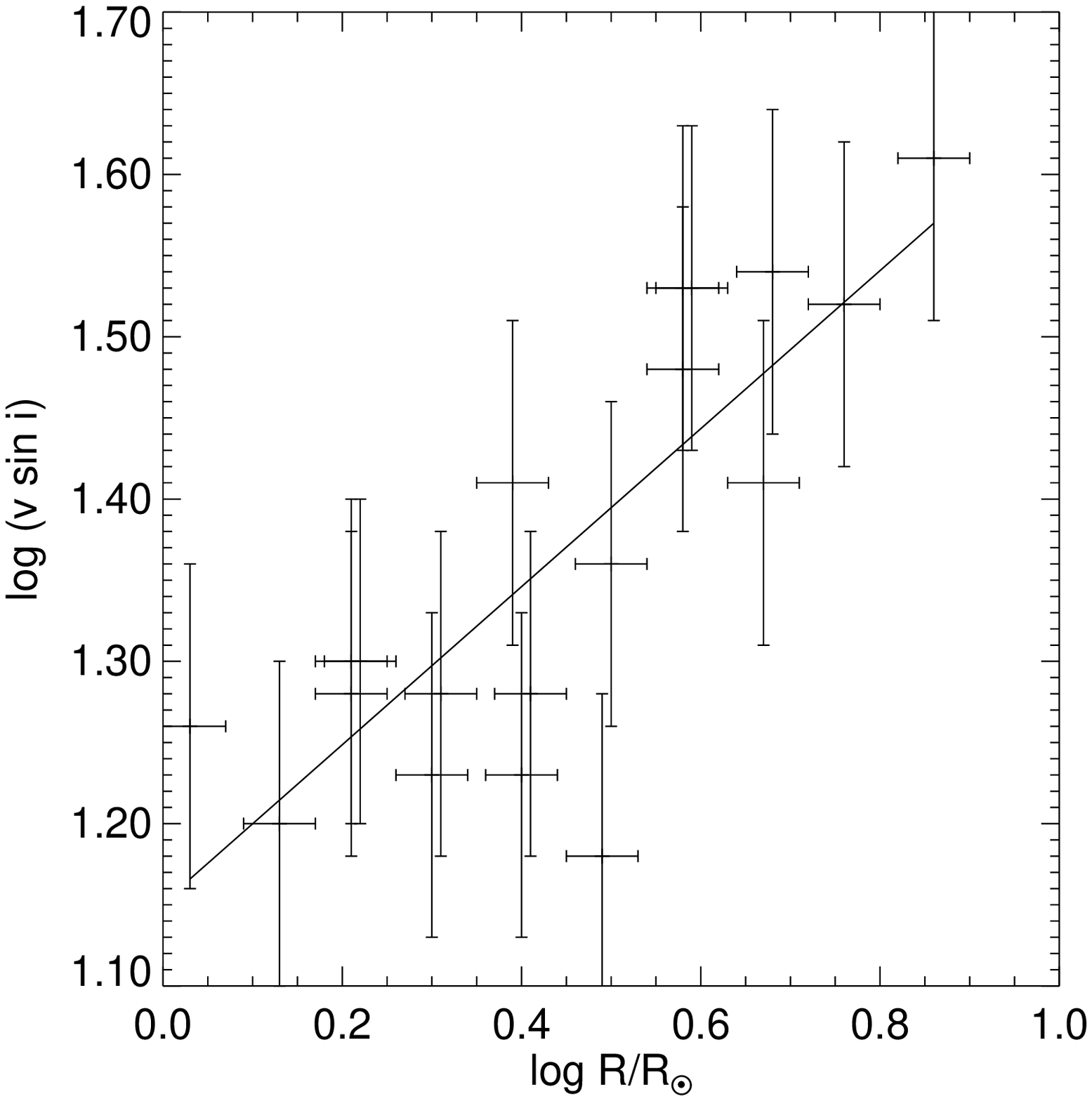}
\caption{Mean values of \vsini\ and log $R$ for each of the (\teff,\lbol)
`boxes' from the previous plots.  Note from previous figures that
$<$\vsini$>$ is nearly constant along lines of constant $R$, thus
encouraging us to combine the data for stars of all masses spanned by the
RHM sample.  The linear fit indicated here is
$\log <v\sin i> = 0.49 (\pm0.11) \log <R> + 1.15 (\pm0.05)$.}
\label{fig:vsinir}
\end{figure*}

These theoretical ideas combined with recent estimates for the lifetimes
of disks may explain why we have not yet detected evidence for spin up in
the samples we have observed.  In particular, Haisch, Lada, \& Lada (2001)
show that more than 80\% of the stars in NGC 2024, which is about 10$^6$
years old have IR excesses indicative of disks, while only 50\% of the
stars in NGC 2264, which they estimate to have an mean age of
3$\times$10$^6$ years, have such excesses. Assume that {\t all} stars with
ages 10$^6$ years and less are surrounded by disks and that during this
time, their rotation periods are locked. Suppose further that half of
these stars lose their disks sometime between 10$^6$ and 3$\times$10$^6$
years.  Reference to Figures \ref{fig:pvsrff}, \ref{fig:pvsr2264}, and
\ref{fig:pvsronc} shows that fully half of our sample stars have ages
$\sim$1 Myr or less and thus are, by hypothesis, locked to their disks.
The remaining stars in our sample have ages between $\sim$1 and $\sim$3
Myr. Of those stars, at least half should still be surrounded by disks.
Hence less than a quarter of our sample might be free to spin up, and most
of these objects are likely to have lost their disks within the past 1
Myr, and not have spun up significantly. If these estimates of disk
lifetimes are correct, it seems plausible that for the distribution of
ages represented among our sample stars and the likely distribution of
disk lifetimes, that such modest spinup might well be masked by the
intrinsic spread in $P$ characteristic of the youngest (presumably
disk-locked) stars in our sample.

The essential point, however, is that both the observations reported here
and theory seem to require that a braking mechanism--whatever it may
be--must be effective for most pre-main sequence stars during at least the
first 3 Myr or so after they reach the birthline.

Significant spin-up must ultimately occur in at least some stars in order
to account for the rapid rotators seen on the main sequence in young
clusters such as Alpha Persei (Stauffer \etal\ 1989).  However, recent
observations of $v \sin i$ for a small sample of stars in the 10 Myr old
TW Hya association (Torres \etal\ 2001; Sterzik \etal\ 1999) suggest mean
$v \sin i$ values similar to those found among the oldest stars in our NGC
2264 sample, despite the fact that the nominal radii among the TW Hya
stars are smaller by 50\%.  These observations suggest that ``$\omega$
locking" may extend to 10 Myr and makes urgent the need for a campaign
focused on mapping the angular momentum evolution of low mass stars in the
age range 3$-$30 Myr so as to understand when rotational spinup in
response to contraction takes place.  Additionally, it would help to have
an age indicator other than radius or else to show that radius correlates
at least roughly with independent age estimates for clusters other than
the ONC.

\begin{acknowledgements}
We wish to thank Jonathan Lunine and Lee Hartmann for several helpful
discussions regarding possible explanations for the apparent paradox
explored here.  We wish to thank our referee Bill Herbst for 
extensive comments on the original manuscript.
SES acknowledges support from the NASA Origins of Solar
Systems program which enabled analysis of the NGC 2264 data.  We thank
Mark Adams for multiple comments and support during the early phases of
the investigation of periodic stars and the McDonald Observatory for the
award of guest investigator time on the 0.9m telescope.  The research
described in this paper was partially carried out at the Jet Propulsion
Laboratory, California Institute of Technology, under a contract with the
National Aeronautics and Space Administration.
\end{acknowledgements}

\begin{table*}[tbp]
\caption{Surveys of stars in the Orion region \label{tab:otherstudies}}
\begin{flushleft}
\begin{tabular}{lccccc} \hline
description     & SMMV\tablenotemark{a} & HRHC\tablenotemark{a} &
R01\tablenotemark{a} & CHS\tablenotemark{a} & HBJM\tablenotemark{a} \\
\hline
\# stars monitored & 2279 & 829 & 3585 & 17,808 & 2294 \\
monitoring band & \ic\ & \ic\ & \ic\ & $JHK$ & \ic\ \\
\# periods reported & 254 & 134 & 281 & 233 & 369\tablenotemark{b} \\
range of sensitivity (d) & 0.1$-$10 & 1$-$35?\tablenotemark{c} & 0.2$-$40 & 1$-$13 &
0.2$-$30?\tablenotemark{c}\\
range of completeness (d) & 0.1$-$6 & 1$-$?\tablenotemark{c} & 0.3$-$20 &
2$-$10 & 0.8$-$15?\tablenotemark{c}\\
approx.\ mass range (\msun)& 0.15$-$1.0 & 0.2$-$2.0 & 0.2$-$1.3 & 0.2$-$1.0 & 0.1$-$1
\\
\# stars used here\tablenotemark{d} & 142 (93) & 95 (30) & 220 (173) & 70
(16) & 211 (145) \\
\hline
\end{tabular}
\tablenotetext{a}{SMMV=Stassun \etal\ (1999), HRHC=Herbst \etal\ (2000),
R01=Rebull (2001), CHS=Carpenter \etal\ (2001), HBHM = Herbst \etal\
(2001)}
\tablenotetext{b}{404 stars were reported in HBJM, but 369 remain in the
most current list provided by W. Herbst.}
\tablenotetext{c}{The range over which this study is complete is not
reported, and completeness cannot be determined without specific times
of observations, which are also not reported. }
\tablenotetext{d}{The most significant limitation on the databases is that
we require that spectral types and $VI_{\rm C}K_s$ photometry be available
for all the stars included in the present study.  However, because of
significant overlap in each of these Orion fields, there are of course
duplicate stars in these databases.  The second number in parentheses in
each column denotes the number of stars actually used here from the
corresponding database.  For example, there are 142 stars in the SMMV
database for which SMMV measured a period, but only 93 for which they
alone measured a period (or have the best period available).  In the
remaining 49 stars, another more recent study retrieved an identical
period, or has made a convincing case that it has the better period
for those stars.}
\end{flushleft}
\end{table*}

\begin{deluxetable}{cccccccccccc}
\tablecolumns{5}
\tablewidth{0pc}
\tablecaption{Mean magnitude of empirically-derived errors
\label{tab:errors}}
\tablehead{
\colhead{category} &
\colhead{uncertainty} &
\colhead{$<$K5} &
\colhead{K5$-$M2} &
\colhead{$>$M2} }
\startdata
photometric & $\delta$log \teff\ & 0.004 & 0.002 & 0.002 \\
in $V-I$    & $\delta$log \lbol\ & 0.052 & 0.044 & 0.039 \\
(\S\ref{sect:photerr})
            & $\delta$log  $R$	& 0.027 & 0.022 & 0.020 \\
\hline
photometric & $\delta$log \teff\ & 0     & 0     & 0     \\
in $I$ +    & $\delta$log \lbol\ & 0.017 & 0.025 & 0.022 \\
amplitude (\S\ref{sect:photerr})
            & $\delta$log  $R$   & 0.009 & 0.013 & 0.011 \\
\hline
spec type   & $\delta$log \teff\ & 0.013 & 0.012 & 0.010  \\
(\S\ref{sect:sptyerr})
            & $\delta$log \lbol\ & 0.047 & 0.063 & 0.119  \\
            & $\delta$log  $R$   & 0.035 & 0.040 & 0.063  \\
\hline
net mean    & $\delta$log \teff\ & 0.014 & 0.012 & 0.010 \\
            & $\delta$log \lbol\ & 0.072 & 0.081 & 0.127  \\
            & $\delta$log  $R$   & 0.050 & 0.050 & 0.070 \\
\hline
typical     & $\delta$log \teff\ & & 0.01\\
	    & $\delta$log \lbol\ & & 0.08\\
	    & $\delta$log  $R$   & & 0.05\\
\enddata
\end{deluxetable}

\begin{deluxetable}{cccccccccccc}
\tablecolumns{7}
\tablewidth{0pc}
\tablecaption{Mean log $P$ and log $R$ for upper and lower quartiles in log
$R$ (see text)\label{tab:quartiles}}
\tablehead{
\colhead{sample} &
\colhead{upper $< \log P >$} &
\colhead{upper $< \log R >$} &
\colhead{lower $< \log P >$} &
\colhead{lower $< \log R >$} &
\colhead{KS prob\tablenotemark{a}} &
\colhead{$\chi^2$ prob\tablenotemark{a}}
}
\startdata
FF & 0.50$\pm$0.06 & 0.47$\pm$0.01 & 0.58$\pm$0.05 & 0.13$\pm$0.01 & 0.22 &
0.74\\
ONC & 0.61$\pm$0.04 & 0.54$\pm$0.01 & 0.57$\pm$0.04 & 0.15$\pm$0.01 & 0.54
& 0.63\\
FF disks\tablenotemark{b} &  0.47$\pm$0.09 & 0.53$\pm$0.02 & 0.62$\pm$0.09 & 0.16$\pm$0.03 &
0.07 & 0.20 \\
FF no disks\tablenotemark{c} & 0.58$\pm$0.10 & 0.41$\pm$0.02 & 0.46$\pm$0.09 & 0.10$\pm$0.02
& 0.30 & 0.37\\
\enddata
\tablenotetext{a}{The probability derived from a two-sided K-S test of the
continuous (or $\chi^2$ test of the binned) distributions of log $P$ for the
upper quartile of log $R$ compared with that for the lower quartile of log
$R$.  Neither the $\chi^2$ nor the KS test suggest any statistically
significant difference between the upper and lower quartiles. }
\tablenotetext{b}{Those stars with \ik\ excesses$>$0.3 mag}
\tablenotetext{c}{Those stars with \ik\ excesses$<$0.1 mag}
\end{deluxetable}

\begin{deluxetable}{cccccccccccc}
\tablecolumns{12}
\tablewidth{0pc}
\tablecaption{Distribution of $v \sin i$ for the ONC stars. \label{tab:vsini}}
\tablehead{
\colhead{log \teff\tablenotemark{a} $\rightarrow$} &
\multicolumn{2}{c}{A (3.730-3.682)} &
\multicolumn{2}{c}{B (3.682-3.634)} &
\multicolumn{2}{c}{C (3.634-3.586)} &
\multicolumn{2}{c}{D (3.586-3.538)} &
\multicolumn{2}{c}{E (3.538-3.490)} \\
\colhead{log $L/L_{\sun}$\tablenotemark{a}$\downarrow$} &
\colhead{$<$\vsini$>$} & \colhead{num\tablenotemark{b} } &
\colhead{$<$\vsini$>$} & \colhead{num } &
\colhead{$<$\vsini$>$} & \colhead{num } &
\colhead{$<$\vsini$>$} & \colhead{num } &
\colhead{$<$\vsini$>$} & \colhead{num } & \colhead{}
}
\startdata
 1 (1.48 - 1.11)    & 33$\pm$13& 3& 41$\pm$10& 5 \\
 2 (1.11 - 0.74)    & 30$\pm$4 & 6& 26$\pm$6 & 4 & 14 & 1\\
 3 (0.74 - 0.37)    & 26$\pm$5 & 7& 15$\pm$4 & 4 & 34$\pm$12& 7  & 35$\pm$9& 5 \\
 4 (0.37 - 0.00)    & 19$\pm$7 & 2& 17$\pm$3 & 13& 17$\pm$3 & 18 & 23$\pm$4& 17 & 34$\pm$15& 3 \\
 5 (0.00 - $-$0.37)  & 12	& 1&	    &	& 20$\pm$8 & 6  & 19$\pm$3& 41 & 19$\pm$2 & 22\\
 6 ($-$0.37 - $-$0.74)& 	&  &	       &   & 18$\pm$7 & 3  & 16$\pm$4& 18 & 20$\pm$3 & 44\\
\enddata
\tablenotetext{a}{Subdivisions defined exactly as in RHM; subdivisions
according to \teff\ and $L$ correspond to subdivisions by $M$ for stars
on convective tracks, and with age for stars of a given $M$. Column A from
RHM spans a larger range of masses than the
other columns, which span log $M/M_{\sun}\sim$0.1 dex when calculated via
D'Antona \& Mazzitelli (1994) or Siess \etal\ (2000).}
\tablenotetext{b}{Number of stars in the bin.}
\end{deluxetable}

\begin{deluxetable}{cccccccccccc}
\tablecolumns{12}
\tablewidth{0pc}
\tablecaption{Trends in angular momentum ($\propto vR$) and angular
velocity ($\propto v/R$) for the ONC.  \label{tab:vrvr}}
\tablehead{
\colhead{log \teff\tablenotemark{a} $\rightarrow$} &
\multicolumn{2}{c}{A (3.730-3.682)} &
\multicolumn{2}{c}{B (3.682-3.634)} &
\multicolumn{2}{c}{C (3.634-3.586)} &
\multicolumn{2}{c}{D (3.586-3.538)} &
\multicolumn{2}{c}{E (3.538-3.490)} \\
\colhead{log $L/L_{\sun}$\tablenotemark{a}$\downarrow$} &
\colhead{$<v \sin i>/R$\tablenotemark{b}} &
\colhead{$<v \sin i>\times R$\tablenotemark{b}} &
\colhead{$<v \sin i>/R$} & \colhead{$<v \sin i>\times R$} &
\colhead{$<v \sin i>/R$} & \colhead{$<v \sin i>\times R$} &
\colhead{$<v \sin i>/R$} & \colhead{$<v \sin i>\times R$} &
\colhead{$<v \sin i>/R$} & \colhead{$<v \sin i>\times R$} }
\startdata
 1 (1.48 - 1.11)    & 6  &191 &  6 &296 &	\\
 2 (1.11 - 0.74)    & 8  &112 &  6 &123 & 2  &  82  \\
 3 (0.74 - 0.37)    & 11 &64  &  5 &46  & 9  &  131 & 7 & 168\\
 4 (0.37 - 0.00)    & 12 &31  &  8 &34  & 7  &  43  & 7 & 72 & 9 & 133 \\
 5 (0.00 - $-$0.37)  & 11 &13  &    &  & 12 &  33  & 9 & 39 & 7 & 48  \\
 6 ($-$0.37 - $-$0.74)&    &	&    & & 17 &  19  & 12& 21 & 12& 33  \\
\enddata
\tablenotetext{a}{Subdivisions defined exactly as in RHM; subdivisions
according to \teff\ and $L$ correspond to subdivisions by $M$ for stars
on convective tracks, and with age for stars of a given $M$. Column A from
RHM spans a larger range of masses than the
other columns, which span log $M/M_{\sun}\sim$0.1 dex when calculated via
D'Antona \& Mazzitelli (1994) or Siess \etal\ (2000).}
\tablenotetext{b}{$<v \sin i>/R$ would be constant if stellar $J$ is
conserved,  and $<v \sin i>/R$ would be constant if $\omega$ conserved.
Note that $<v \sin i>R$ changes, whereas $<v \sin i>/R$ is constant to
within a factor of $\sim$2.}
\end{deluxetable}

\begin{thebibliography}{dummy}
\bibitem[1]{1}Barnes, S.\ A.\ 2001, ApJ, 561, 1095
\bibitem[Bessell(1991)]{bessell}Bessell, M.\ S.\ 1991, \aj, 101, 662
\bibitem[1.5]{1.5}Bevington, P.\ R., \& Robinson, D.\ K.\ 1992, Data
Reduction and Error Analysis for the Physical Sciences (2d ed; New York:
McGraw-Hill)
\bibitem[2]{2}Bouvier, J., Forestini, M., \& Allain, S.\ 1997, \aap, 326, 1023
\bibitem[CHS]{chs} Carpenter, J.\ M., Hillenbrand, L.\
A., \& Skrutskie, M., F.\ 2001, AJ, 121, 3160 (CHS)
\bibitem[3]{3}Choi, P., \& Herbst, W.\ 1996, AJ, 111, 283
\bibitem[4]{4}D'Antona, F., \& Mazzitelli, I., 1994, \apjs, 90, 467
\bibitem[5]{5}Durisen, R.\ H., Yang, S., Cassen, P., \& Stahler, S.\ 1989,
\apj, 345, 959
\bibitem[6]{edwards93}Edwards, S., Strom, S.\ E., Hartigan, P., Strom,
K.\ M., Hillenbrand, L.\ A., Herbst, W., Attridge, J., Meriill, K.\
M., Probst, R., \& Gatley, I., 1993, \aj, 106, 372
\bibitem[7]{7}Genzel, R., Reid, J.\ J., Moran, J.\ M., \& Downes, D.\ 1981,
\apj, 244, 884
\bibitem[8]{8}Haisch, K., Lada, E., \& Lada, C.\ 2001, \apjl, 553, 153
\bibitem[9]{9}Hartigan, P., Strom, K., \& Strom, S., 1994, ApJ, 427, 961
\bibitem[10]{10}Hartmann, L.\ 2001, AJ, 121, 1030
\bibitem[11]{11}Hartmann, L., Cassen, P., \& Kenyon, S.\ J.\ 1997, \apj, 475, 770
\bibitem[12]{12}Herbig, G.\ 1954, ApJ, 119, 483
\bibitem[13]{13}Herbst, W., Bailer-Jones, C.\ A.\ L., \& Mundt, R. 2001,
\apjl, 554, 197 (HBJM)
\bibitem[14]{14}Herbst, W., Rhode, K., Hillenbrand, L., \& Curran, G.\
2000, \aj, 119, 261 (HRHC)
\bibitem[15]{15}Hillenbrand, L.\ A.\ 1997, \aj, 113, 1733
\bibitem[16]{16}Hillenbrand, L.\ A., \etal\ 1998, \aj, 116, 1816
\bibitem[17]{17}Kawaler, S.\ D.\ 1988, \apj, 333, 236
\bibitem[18]{18}Kearns, K., \& Herbst, W., 1997, AJ, 114, 1098
\bibitem[19]{19}Kearns, K., \& Herbst, W., 1998, AJ, 116, 261
\bibitem[20]{20}K\"onigl, A.\ 1991, \apj, 370, 39
\bibitem[Leggett(1992)]{leggett92}Leggett, S.\ K.\ 1992, \apjs, 82, 351
\bibitem[Leggett(1998)]{leggett98}Leggett, S.\ K., Allard, F., Hauschildt,
P.\ H.\ 1998, \apj, 509, 836
\bibitem[20.5]{20.5}Lada, C.\ J., \etal\ 2000, AJ, 120, 3162
\bibitem[21]{21}MacGregor, K.\ B., \& Charbonneau, P.\ 1994, in ASP Conf.\
Ser.\ 64, Cool Stars, Stellar Systems, and the Sun, Eighth Cambridge
Workshop, ed.\ J.-P.\ Caillaut (San Francisco:ASP), 174
\bibitem[22]{22}Makidon, R., Rebull, L., Strom, S., Adams, M., \& Patten, B.
2002, in preparation
\bibitem[23]{23}Palla, F., \& Stahler, S.\ W.\ 1992, \apj, 392, 667
\bibitem[25]{25}Rebull, L.\ M.\ 2001, \aj, 121, 1676
\bibitem[26]{26}Rebull, L.\ M., Hillenbrand, L., Strom S.\ E., Duncan, D.\ K.,
Patten, B.\ M., Pavlovsky, C.\ M., Makidon, R., \& Adams, M.\ T.\ 2000, \aj,
119, 3026
\bibitem[27]{27}Rebull, L.\ M., Makidon, R.\ B., Strom, S.\ E., Hillenbrand,
L.\ A., Birmingham, A., Patten, B.\ M., Jones, B.\ F., Yagi, H., \&
Adams, M.\ T.\ 2002, AJ, in press
\bibitem[28]{28}Rhode, K.\ L., Herbst, W., \& Mathieu, R.\ D. 2001, \aj, 
122, 3258 (RHM)
\bibitem[29]{29}Soderblom, D., \etal\ 1999, AJ, 118, 1301
\bibitem[30]{30}Shu, F., Najita, J., Shang, H., \& Li, Z.-Y.\ 2000, in
Protostars and Planets IV, ed.\ V.\ Mannings, A.\ P.\ Boss \& S.\ S.\
Russell (Tucson: University of Arizona Press), 789
\bibitem[31]{31}Siess, L., Dufour, E., \& Forestini, M., 2000, \aap, 358, 593
\bibitem[32]{32}Stassun, K.\ G., Mathieu, R.\ D., Mazeh, T., \& Vrba, F.\ J.\
1999, AJ, 117, 2941 (SMMV)
\bibitem[32.5]{32.5}Stassun, K.\ G., Mathieu, R.\ D., Vrba, F.\ J.,
Mazeh, T., Henden, A.\ 2001, AJ, 121, 1003
\bibitem[33]{33}Stauffer, J.\ R., Hartmann, L.\ W., \& Jones, B.\ F.\ 1989,
\apj, 346, 160
\bibitem[34]{34}Sterzik, M., Alcal\'a, J., Covino, E., Petr, M.\ 1999, A\&A,
346, L41
\bibitem[35]{35}Sung, H., Bessell, M.\ S., \& Lee, S-W.\ 1997, \aj, 114, 2644
\bibitem[36]{36}Swenson, F.\ J., Faulkner, J., Rogers, F.\ J., \& Iglesias,
C.\ A.\ 1994, \apj, 425, 286
\bibitem[37]{37}Tinker, J., Pinsonneault, M., \& Terndrup, D.\ 2002,
ApJ, 564, 877
\bibitem[38]{38}Torres, G., Neuhauser, R., \& Latham, D., 2001,
astro-ph/0105132, to appear in Young Stars Near Earth: Progress and
Prospects, 2001, Jayawardhana, R., \& Greene, T., eds.
\bibitem[39]{39}Trilling, D.\ E., Benz, W., Guillot, T., Lunine, J.\ I.,
Hubbard, W.\ B., \& Burrows, A.\ 1998, \apj, 500, 428
\end{thebibliography}
\end{document}